\documentclass[twoside,twocolumn,9pt]{article}
\usepackage{extsizes}
\usepackage[super,sort&compress,comma]{natbib} 
\usepackage[version=3]{mhchem}
\usepackage[left=1.5cm, right=1.5cm, top=1.785cm, bottom=2.0cm]{geometry}
\usepackage{balance}
\usepackage{mathptmx}
\usepackage{sectsty}
\usepackage{graphicx} 
\usepackage{lastpage}
\usepackage[format=plain,justification=justified,singlelinecheck=false,font={stretch=1.125,small,sf},labelfont=bf,labelsep=space]{caption}
\usepackage{float}
\usepackage{fancyhdr}
\usepackage{fnpos}
\usepackage[english]{babel}
\addto{\captionsenglish}{%
  
}
\usepackage{array}
\usepackage{droidsans}
\usepackage{charter}
\usepackage[T1]{fontenc}
\usepackage[usenames,dvipsnames]{xcolor}
\usepackage{setspace}
\usepackage[compact]{titlesec}
\usepackage{hyperref}
\usepackage{gensymb}

\usepackage{epstopdf}

\definecolor{cream}{RGB}{222,217,201}

\begin{document}

\pagestyle{fancy}
\thispagestyle{plain}
\fancypagestyle{plain}{
\renewcommand{\headrulewidth}{0pt}
}

\makeFNbottom
\makeatletter
\renewcommand\LARGE{\@setfontsize\LARGE{15pt}{17}}
\renewcommand\Large{\@setfontsize\Large{12pt}{14}}
\renewcommand\large{\@setfontsize\large{10pt}{12}}
\renewcommand\footnotesize{\@setfontsize\footnotesize{7pt}{10}}
\makeatother

\renewcommand{\thefootnote}{\fnsymbol{footnote}}
\renewcommand\footnoterule{\vspace*{1pt}%
\color{cream}\hrule width 3.5in height 0.4pt \color{black}\vspace*{5pt}} 
\setcounter{secnumdepth}{5}

\makeatletter 
\renewcommand\@biblabel[1]{#1}            
\renewcommand\@makefntext[1]%
{\noindent\makebox[0pt][r]{\@thefnmark\,}#1}
\makeatother 
\renewcommand{\figurename}{\small{Fig.}~}
\sectionfont{\sffamily\Large}
\subsectionfont{\normalsize}
\subsubsectionfont{\bf}
\setstretch{1.125} 
\setlength{\skip\footins}{0.8cm}
\setlength{\footnotesep}{0.25cm}
\setlength{\jot}{10pt}
\titlespacing*{\section}{0pt}{4pt}{4pt}
\titlespacing*{\subsection}{0pt}{15pt}{1pt}

\fancyfoot{}
\fancyfoot[LO,RE]{\vspace{-7.1pt}\includegraphics[height=9pt]{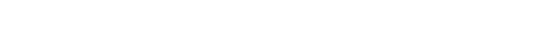}}
\fancyfoot[CO]{\vspace{-7.1pt}\hspace{13.2cm}\includegraphics{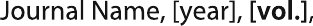}}
\fancyfoot[CE]{\vspace{-7.2pt}\hspace{-14.2cm}\includegraphics{head_foot/RF}}
\fancyfoot[RO]{\footnotesize{\sffamily{1--\pageref{LastPage} ~\textbar  \hspace{2pt}\thepage}}}
\fancyfoot[LE]{\footnotesize{\sffamily{\thepage~\textbar\hspace{3.45cm} 1--\pageref{LastPage}}}}
\fancyhead{}
\renewcommand{\headrulewidth}{0pt} 
\renewcommand{\footrulewidth}{0pt}
\setlength{\arrayrulewidth}{1pt}
\setlength{\columnsep}{6.5mm}
\setlength\bibsep{1pt}

\makeatletter 
\newlength{\figrulesep} 
\setlength{\figrulesep}{0.5\textfloatsep} 

\newcommand{\topfigrule}{\vspace*{-1pt}%
\noindent{\color{cream}\rule[-\figrulesep]{\columnwidth}{1.5pt}} }

\newcommand{\botfigrule}{\vspace*{-2pt}%
\noindent{\color{cream}\rule[\figrulesep]{\columnwidth}{1.5pt}} }

\newcommand{\dblfigrule}{\vspace*{-1pt}%
\noindent{\color{cream}\rule[-\figrulesep]{\textwidth}{1.5pt}} }

\makeatother

\twocolumn[
  \begin{@twocolumnfalse}
{\includegraphics[height=30pt]{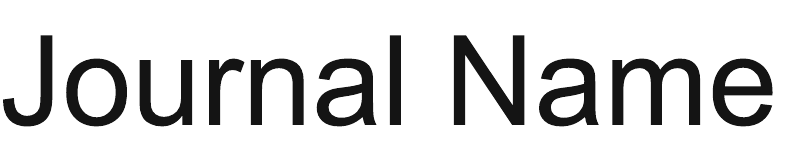}\hfill\raisebox{0pt}[0pt][0pt]{\includegraphics[height=55pt]{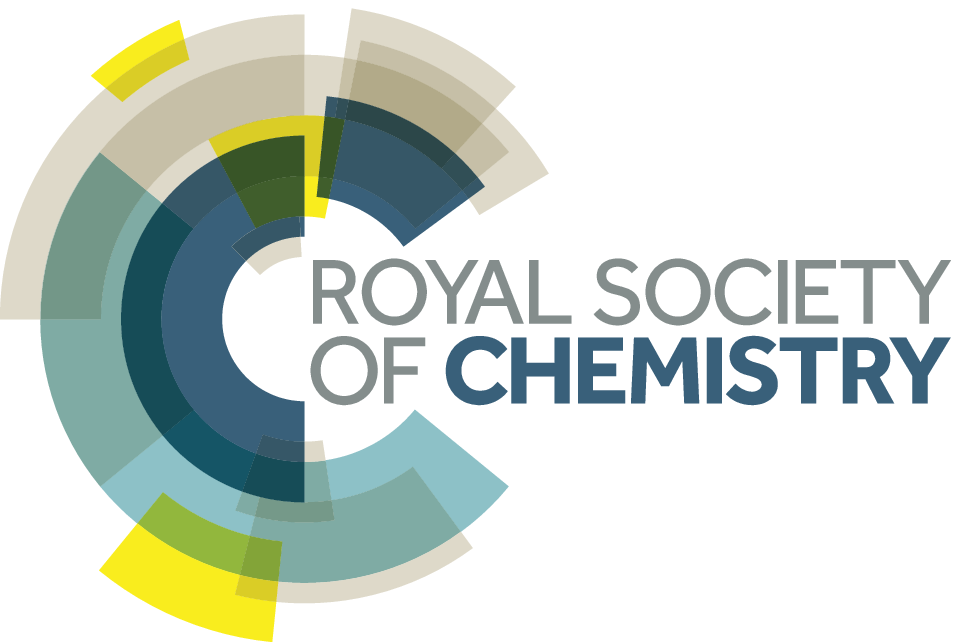}}\\[1ex]
\includegraphics[width=18.5cm]{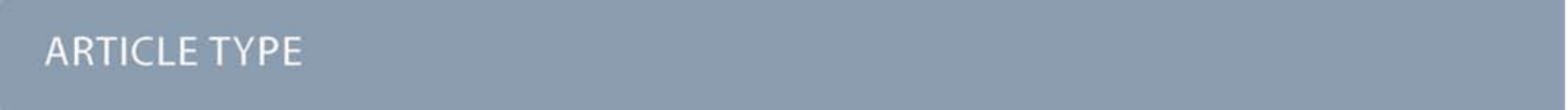}}\par
\vspace{1em}
\sffamily
\begin{tabular}{m{4.5cm} p{13.5cm} }

\includegraphics{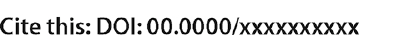} & \noindent\LARGE{\textbf{\textit{In situ} Bragg coherent X-ray diffraction imaging of corrosion in a Co-Fe alloy microcrystal$^\dag$}} \\

\vspace{0.3cm} & \vspace{0.3cm} \\

 & \noindent\large{David Yang,$^{\ast}$\textit{$^{a}$} Nicholas W. Phillips,\textit{$^{a\ddag}$} Kay Song,\textit{$^{a}$} Clara Barker,\textit{$^{b}$} Ross J. Harder,\textit{$^{c}$} Wonsuk Cha,\textit{$^{c}$} Wenjun Liu,\textit{$^{c}$} and Felix Hofmann$^{\ast}$\textit{$^{a}$}} \\

\includegraphics{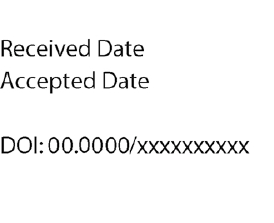} & \noindent\normalsize{Corrosion is a major concern for many industries, as corrosive environments can induce structural and morphological changes that lead to material dissolution and accelerate material failure. The progression of corrosion depends on nanoscale morphology, stress, and defects present. Experimentally monitoring this complex interplay is challenging. Here we implement \textit{in situ} Bragg coherent X-ray diffraction imaging (BCDI) to probe the dissolution of a Co-Fe alloy microcrystal exposed to hydrochloric acid (HCl). By measuring five Bragg reflections from a single isolated microcrystal at ambient conditions, we compare the full three-dimensional (3D) strain state before corrosion and the strain along the $[111]$ direction throughout the corrosion process. We find that the strained surface layer of the crystal dissolves to leave a progressively less strained surface. Interestingly, the average strain closer to the centre of the crystal increases during the corrosion process. We determine the localised corrosion rate from BCDI data, revealing the preferential dissolution of facets more exposed to the acid stream, highlighting an experimental geometry effect. These results bring new perspectives to understanding the interplay between crystal strain, morphology, and corrosion; a prerequisite for the design of more corrosion-resistant materials.} \\

\end{tabular}

 \end{@twocolumnfalse} \vspace{0.6cm}

  ]

\renewcommand*\rmdefault{bch}\normalfont\upshape
\rmfamily
\section*{}
\vspace{-1cm}


\footnotetext{\textit{$^{a}$~Department of Engineering Science, University of Oxford, Oxford, OX1 3PJ, United Kingdom. E-mail: david.yang@eng.ox.ac.uk, felix.hofmann@eng.ox.ac.uk}}
\footnotetext{\textit{$^{b}$~Department of Materials, University of Oxford, Oxford, OX1 3PH, United Kingdom.}}
\footnotetext{\textit{$^{c}$~Advanced Photon Source, Argonne National Laboratory, Argonne, IL 60439, USA.}}

\footnotetext{\dag~Electronic Supplementary Information (ESI) available. See DOI: 00.0000/00000000.}

\footnotetext{\ddag~Present address: Paul Scherrer Institut,
5232 Villigen PSI, Switzerland} 


\section{Introduction}
Corrosion processes are important in a wide variety fields such as geology \cite{Lasaga1998}, renewable energy \cite{Sahin2008}, and catalysis \cite{Strasser2010}. There is a keen interest in the investigation of reaction kinetics over different time scales to better predict how a material transforms in a corrosive environment. However, often the corrosion rates reported in the laboratory setting and in the field are inconsistent, despite  well-controlled extrinsic conditions of the experiment \cite{Fischer2012,Fischer2014,Noiriel2020}. Frequently, the cause for the corrosion rate discrepancy is based on intrinsic factors related to the properties of the sample \cite{Fischer2012,Fischer2014}, for example, morphological features and lattice defects are not accounted for in the bulk dissolution rate determined from mineral powders. Crystals possess heterogeneous spatial reactivity on the crystal surface \cite{Fischer2012,Fischer2014,Lasaga1998}, as reaction rates tend to differ along facets, edges, and vertices \cite{Saldi2017,Noiriel2019}. In addition to these morphological features, dissolution rates can be influenced by internal features such as crystal defects. For example, dislocation-induced etch pits \cite{Cabrera1954,Lasaga1986} are a commonly reported result of the interaction between defect-induced strain fields and corrosion. 
This highlights the importance of studying defect-mediated corrosion and the effects of lattice strain on local nanoscale corrosion. Consolidating our understanding of these processes will aid in the design of more corrosion resistant materials.  

Corrosion mechanisms start with the interaction of the crystal surface with a corrosive medium. Techniques such as interferometry (VSI) \cite{Fischer2007}, atomic-force-microscopy (AFM) \cite{Hillner1992}, digital holographic microscopy \cite{Brand2017}, and phase-shift interferometry \cite{Ueta2013} are used to probe surface dissolution rates. However, these methods primarily focus on measuring upwards-facing sample features at the surface. Liquid cell transmission electron microscopy (LCTEM) techniques have been used to study pitting corrosion \cite{Chee2015}, but these studies focus on thin samples where surface effects are dominant, and are prone to bubble formation due to radiolysis \cite{Grogan2013}. \textit{Ab initio} molecular dynamics simulations (AIMD) have been used to predict the energy barriers of dissolution for crystal edges, corners, and kinks \cite{Chen2014}, but we note that the direct determination of dissolution rate for spatially resolved surface sites remains uncommon in the laboratory. Three-dimensional localised dissolution information has only been recently reported using transmission X-ray microscopy (TXM) \cite{Yuan2019a} and three-dimensional (3D) X-ray microtomography (XMT) \cite{Noiriel2019,Noiriel2020}. Although TXM and XMT provide a spatial resolution on the order of 50 nm, they are limited to probing morphological features.

Crystal lattice defects beneath the crystal surface must also be considered when investigating corrosion. Defects can induce strain in the crystal lattice, which changes the activation barrier for dissolution near the defect \cite{Schott1989}. Areas having dislocations, grain boundaries, and high surface roughness dissolve faster than defect-free regions \cite{Lasaga1998}. Indeed, it was through the appearance of etch pits that dislocations were first visualised \cite{Vogel1956}. Experiments have shown that crystal lattice strain can noticeably increase the dissolution rate for calcite \cite{Schott1989,Gutman1994} and sodium chlorate crystals \cite{Morel2001} compared to theoretical values. A full picture of crystal corrosion mechanisms can therefore only be obtained by also studying how the evolution of defect networks, lattice strains within the crystal and local morphology modify the local corrosion characteristics.

To elucidate how both crystal strain and morphology influence localised corrosion, we use Bragg coherent X-ray diffraction imaging (BCDI) to study the evolution of a Co-Fe alloy microcrystal. A binary alloy was chosen for this study to identify how strain distributions can be affected by the presence of another element during corrosion \cite{Cha2017,Chen-Wiegart2017}. BCDI allows 3D-resolved, nanoscale strain measurements with a 3D spatial resolution of 10 - 30 nm and a strain resolution on the order of $\sim$2$\times10^{-4}$ \cite{Hofmann2017a}, making it well suited to probing crystal defects and morphology. Currently, at third generation synchrotron light sources, several minutes are required to collect a BCDI dataset. At these timescales, BCDI allows the \textit{in situ} probing of slow dynamic changes during the corrosion process. Several \textit{in situ} BCDI experiments have been reported, for example, Pt nanocrystals during methane oxidation \cite{Kim2018d}, silver–gold alloys undergoing nitric acid-induced dealloying \cite{Cha2017,Chen-Wiegart2017}, and magnetite reacting with hydrochloric acid (HCl) \cite{Yuan2019}. 

BCDI involves illuminating a single crystal sample with a coherent X-ray beam. The sample must be sufficiently small to fit within the coherence volume of the beam, which typically restricts the sample dimensions to less than $1 \mathrm{\mu m}$ at third generation synchrotron sources. Once the Bragg condition is met for a specific $hkl$ reflection, the diffraction pattern is collected on a pixelated area detector positioned perpendicular to the outgoing wave vector in the Fraunhofer diffraction regime. By rotating the sample through the Bragg condition, a 3D coherent X-ray diffraction pattern (CXDP) is collected as different parts of the 3D Bragg peak sequentially intersect the Ewald sphere in reciprocal space, which is projected onto the detector.

The intensity of the diffraction data is proportional to the magnitude of the Fourier transform, $\mathcal{F}$, of the complex electron density, $\mathbf{\rho(r)}$, where $\mathbf{r}$ is the position vector, of the crystalline volume for a particular crystal reflection. The phase, $\mathbf{\psi(r)}$, corresponds to the projection of the lattice displacement field, $\mathbf{u(r)}$, onto the Bragg vector, $\mathbf{Q_\mathit{hkl}}$, of the crystal reflection under consideration: 

\begin{equation} \label{eq:phase}
    \mathbf{\psi_\mathit{hkl}(r)} = \mathbf{Q_\mathit{hkl}}\cdot\mathbf{u(r)}.
\end{equation}

However, the phase is lost because we only record the square of the amplitude in the intensity, $I_{hkl}$ of the CXDP, as $I_{hkl} \propto |\mathcal{F}(\rho(\mathbf{r})e^{i\psi_{hkl}(\mathbf{r})})|^2$. In order to facilitate recovery of the phase, the CXDP must be oversampled by at least twice the Nyquist frequency \cite{Miao2000b}, at least 4 pixels per fringe period. Iterative phase retrieval algorithms that apply constraints in real and reciprocal space are then used to recover the phase \cite{Robinson2001}. The result is transformed from detector conjugated space to orthogonal lab or sample space \cite{Yang2019}.

Here, we image the spatial and temporal evolution of a Co-Fe alloy microcrystal submerged in an aqueous HCl solution (pH = 0.65) using \textit{in situ} BCDI. Prior to corrosion, we measured five crystal reflections to determine the full lattice strain and rotation tensor, providing a clearer understanding of the defects initially present inside the crystal. During the corrosion, we measure the $(111)$ reflection at 2.6, 4 and 6 hours after the start of acid exposure to examine how the 3D strain, crystal morphology, and local variations in dissolution rate evolve as a function of corrosion. The experiment provides a new perspective for the analysis of corrosion, by probing how both strain and morphology affect the localised dissolution of the crystal.

\section{Methods}
\subsection{Sample manufacture}
\begin{figure}[ht] 
    \centering
    \includegraphics[height=11cm]{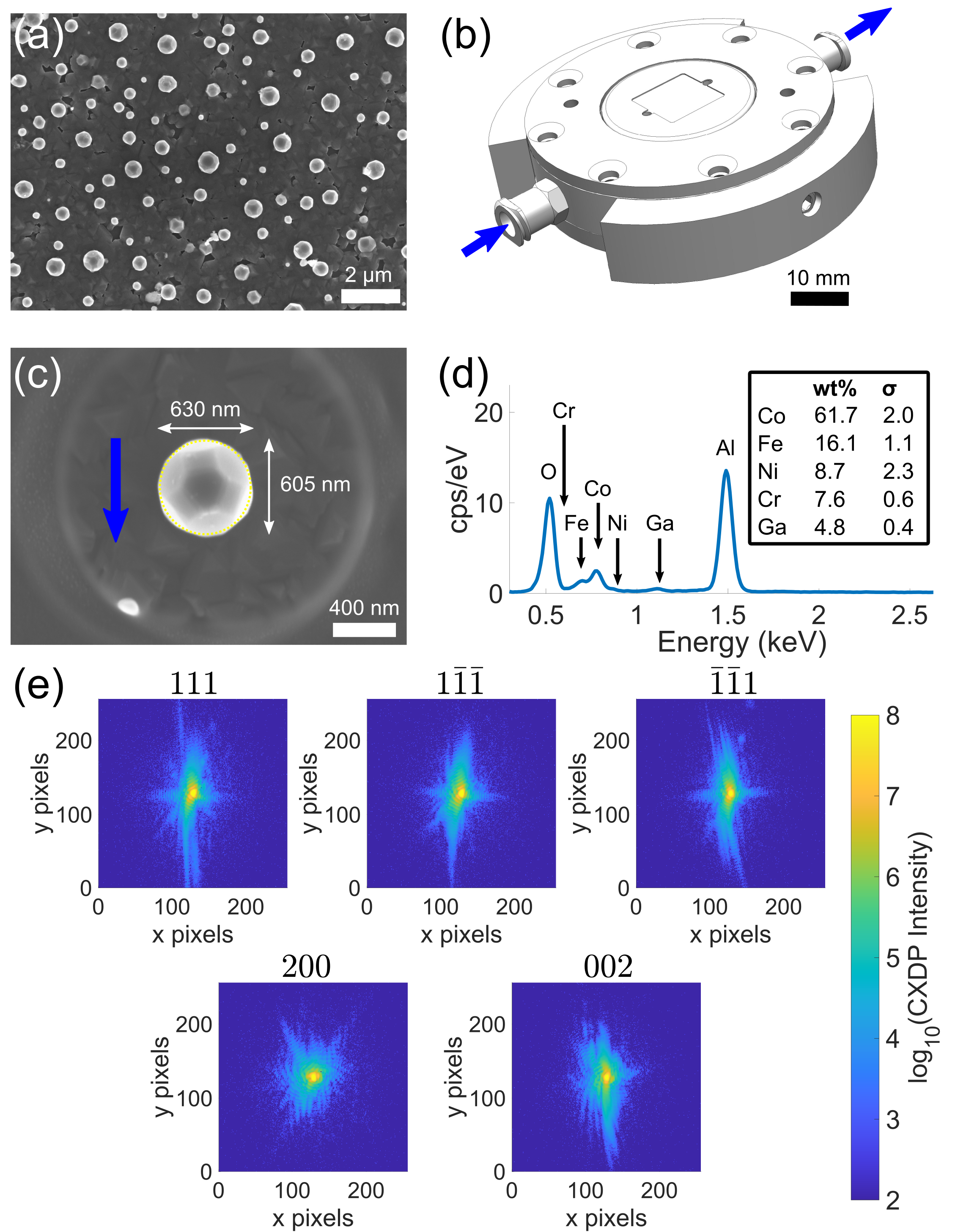}
    \caption{(a) Dewetted Co-Fe alloy microcrystals on sapphire substrate. (b) A schematic of the fluid flow cell at beamline 34-ID-C (Advanced Photon Source) used for \textit{in situ} corrosion. The blue arrows show the direction of fluid flow. (c) The Co-Fe microcrystal measured during the experiment before corrosion. The dotted yellow line represents the region measured using energy-dispersive X-ray spectroscopy (EDX). (d) EDX data for the FIB-isolated sample. Only the L-lines for the most pronounced elements in the crystal are shown above. All elements are homogeneous throughout the crystal (see Appendix \ref{appendix:EDX_results}). The spectrum composition excludes the Al and O substrate peaks. (e) Central slices of the CXDP for each reflection measured before corrosion.}
    \label{fig:SEM_pic}
\end{figure}

Samples were produced by sputter deposition of a thin film onto a single crystal sapphire wafer (C-plane orientation). The thickness of the film was 100 nm. The thin film was dewetted in a vacuum furnace purged with a 5\% hydrogen, balance Argon, gas mixture at 1250°C for 16 hours. The resulting crystals exhibit a face-centred cubic structure, range from 100 nm to 1 $\mathrm{\mu m}$ in size, (Fig. \ref{fig:SEM_pic}(a)) and adhere to the substrate surface.

The samples were coated with 10 nm of amorphous carbon to assist with scanning electron microscope (SEM) imaging. The coating was applied through thermal evaporation using a Leica ACE600 Coater. SEM images were taken on a ZEISS Auriga focused ion beam (FIB)-SEM equipped with an in-lens detector with an acceleration voltage of 7 kV. To facilitate reliable measurement of multiple reflections from a specific microcrystal, Ga ion FIB was used to remove the surrounding crystals within a 20 $\mathrm{\mu m}$ radius using a current of 150 pA - 6 nA. The isolated crystal is shown in Fig. \ref{fig:SEM_pic}(c). Only SEM imaging was used to position the FIB milling scan, as our previous results indicate that even a single low dose FIB image can induce large lattice strains \cite{Hofmann2017a}. Energy-dispersive X-ray spectroscopy (EDX) analysis was used to determine the elemental composition of the crystal (Fig. \ref{fig:SEM_pic}(c)). All elements are homogeneous throughout the crystal (see Appendix \ref{appendix:EDX_results}). EDX was performed on a ZEISS EVO equipped with an X-act detector (Oxford Instruments). The EDX spectrum in Fig \ref{fig:SEM_pic}(d) was acquired on an elliptical region encapsulating the crystal for 104 seconds using an acceleration voltage of 10 kV.

\subsection{Experimental measurements}
Synchrotron X-ray diffraction measurements were performed at the Advanced Photon Source (APS), Argonne National Lab, USA. Prior to BCDI measurements, micro-beam Laue diffraction at beamline 34-ID-E was used to determine the lattice orientation of the crystal. This served to pre-align the crystal for BCDI measurements at beamline 34-ID-C, as previously described \cite{Hofmann2017b}.

The substrate was attached to the fluid flow cell (Fig. \ref{fig:SEM_pic}(b)) provided at beamline 34-ID-C using dental wax. Diffraction measurements used an X-ray energy of 9 keV ($\lambda$ = 0.138 nm), with a bandwith of $\delta\lambda/\lambda \approx 1.3\times10^{-4}$ from a Si$(111)$ monochromator. The X-ray beam was focused to a size of 740 nm × 680 nm (h × v, full width at half-maximum) using Kirkpatrick-Baez (KB) mirrors. Beam defining slits were used to select the coherent portion of the beam at the entrance to the KB mirrors. For beamline 34-ID-C, the transverse coherence is $\xi_h > 10\ \mathrm{\mu m}$ and the longitudinal coherence is $\xi_w\approx0.7\ \mathrm{\mu m}$ at an energy of 9 keV \cite{Leake2009}.

CXDPs were collected on a 256 × 256 pixel Timepix area detector (Amsterdam Scientific Instruments) with a GaAs sensor and pixel size, $p$, of $55 \mathrm{\ \mu m} \times 55 \mathrm{\ \mu m}$ positioned at 0.75 m from the sample to ensure oversampling. CXDPs were recorded by rotating the crystal through an angular range of 0.7° and recording an image every 0.005° with 0.1 s exposure time and 35 accumulations at each angle. The lower bound for the detector distance was determined by $2dp/\lambda = 0.50\ \mathrm{m}$, where $d$ is the sample size.

To optimise the signal to noise ratio and increase the dynamic range of the CXDPs, multiple repeated scans of each reflection were performed at ambient conditions and aligned to maximise their cross-correlation. Once aligned, the minimum acceptable Pearson cross-correlation for summation of CXDPs from a specific Bragg reflection was chosen to be 0.981, similar to other BCDI studies \cite{Hofmann2018,Hofmann2020}. CXDPs were corrected for dead-time, darkfield, and whitefield prior to cross-correlation alignment. CXDPs from the following reflections were collected (the number of repeat scans that were averaged is noted in [] brackets): $(111)$ [10], $(\bar{1}\bar{1}1)$ [9], $(1\bar{1}\bar{1})$ [9], $(200)$ [10], $(002)$ [10]. Details regarding the recovery of the real space images using phase retrieval algorithms can be found in Appendix \ref{appendix:phase_retrieval} and the computation of the strain can be found in Appendix \ref{appendix:strain_calculations}.

\subsection{Corrosion}
The fluid flow cell was covered with a 0.0508 mm thick mylar film. HCl solution (0.22 mol/L, pH=0.65) was injected at a flow rate of 40 mL/min during the full 6 hours. Single scans for the $(111)$ reflection for the crystal were measured at 2.6, 4 and 6 hours after the start of acid injection. CXDPs were recorded and processed using the same procedure as before corrosion.

\section{Results} \label{Results}
\begin{figure}[ht] 
    \centering
    \includegraphics[height=9.5cm]{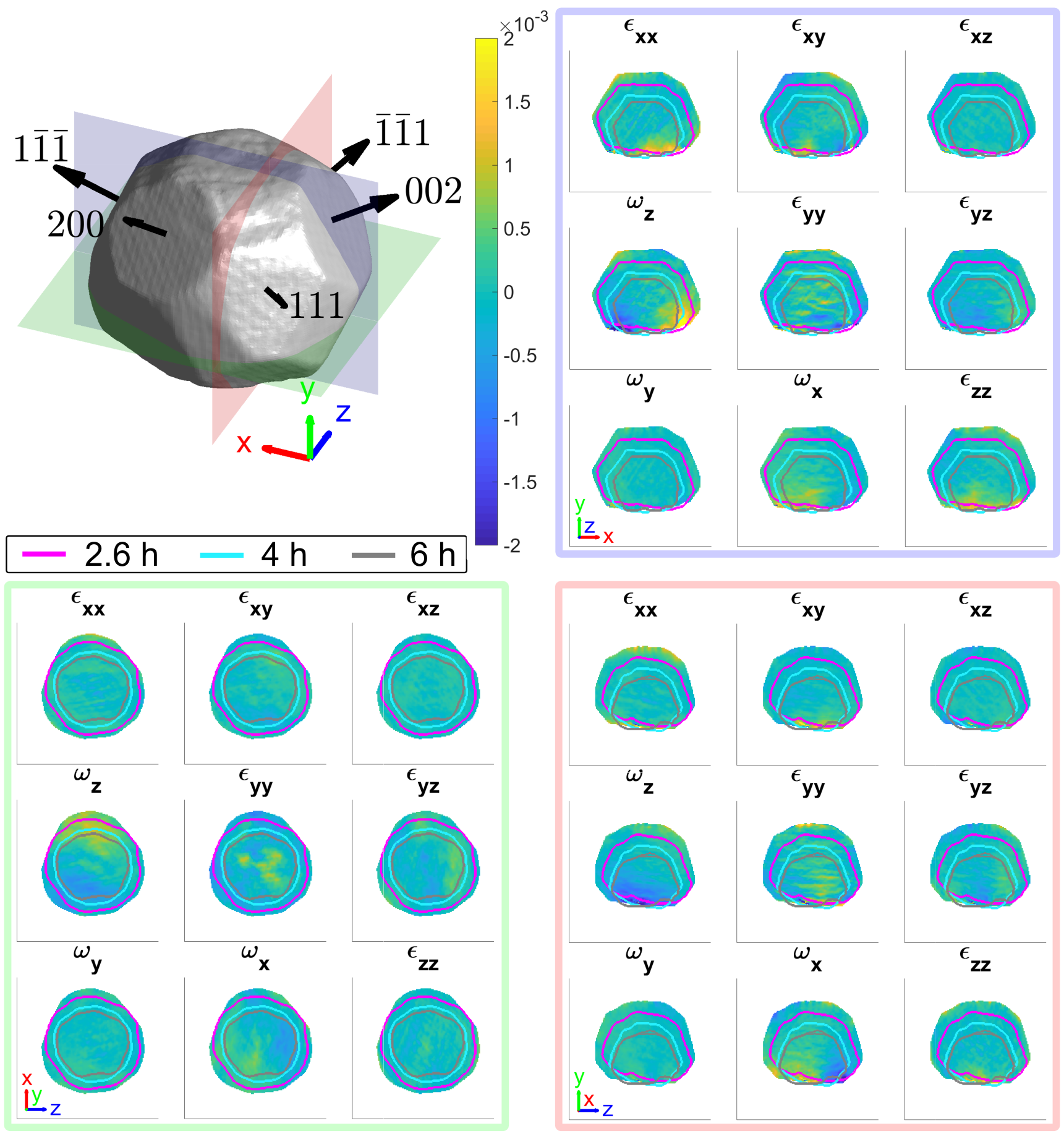}
    \caption{Average morphology of the $(111)$, $(1\bar{1}\bar{1})$, $(\bar{1}\bar{1}1)$, $(200)$, $(002)$ reflections at ambient conditions. The slices through the strain and rotation tensor components are shown for the planes indicated at $x = -87.5\ \mathrm{nm}$ (red), $y = -67.5\ \mathrm{nm}$ (green), and $z = -7.5\ \mathrm{nm}$ (blue) from the centre of mass of the microcrystal. The outlines in the strain and rotation tensor figure panels corresponds to the $(111)$ reconstruction after 2.6 h (magenta), 4 h (cyan) and  6 h (grey) after corrosion. The amplitude threshold is 0.25 and the magnitude of the coordinate axes is 100 nm. Supplementary video (SV) 1-3 show the strain and rotation tensor components throughout the volume along the $x$, $y$, and $z$ axis respectively.}
    \label{fig:Tensor}
\end{figure}

Fig. \ref{fig:Tensor} shows the strain and rotation tensors reconstructed from five measured Bragg reflections. The average amplitude taken over all five reflections is based on an amplitude threshold of 0.25.

The strain and rotation tensor components are fairly homogeneous, similar to other crystals fabricated by dewetting methods \cite{Robinson2001,Hofmann2018,Hofmann2017a}. Numerous small surface defects are visible in the cross sectional edges of the crystal. At the base of each cross section there are regions of strain that are likely due to lattice mismatch with the sapphire substrate \cite{Beutier2013}.  



\begin{figure}[ht] 
    \centering
    \includegraphics[height=8cm]{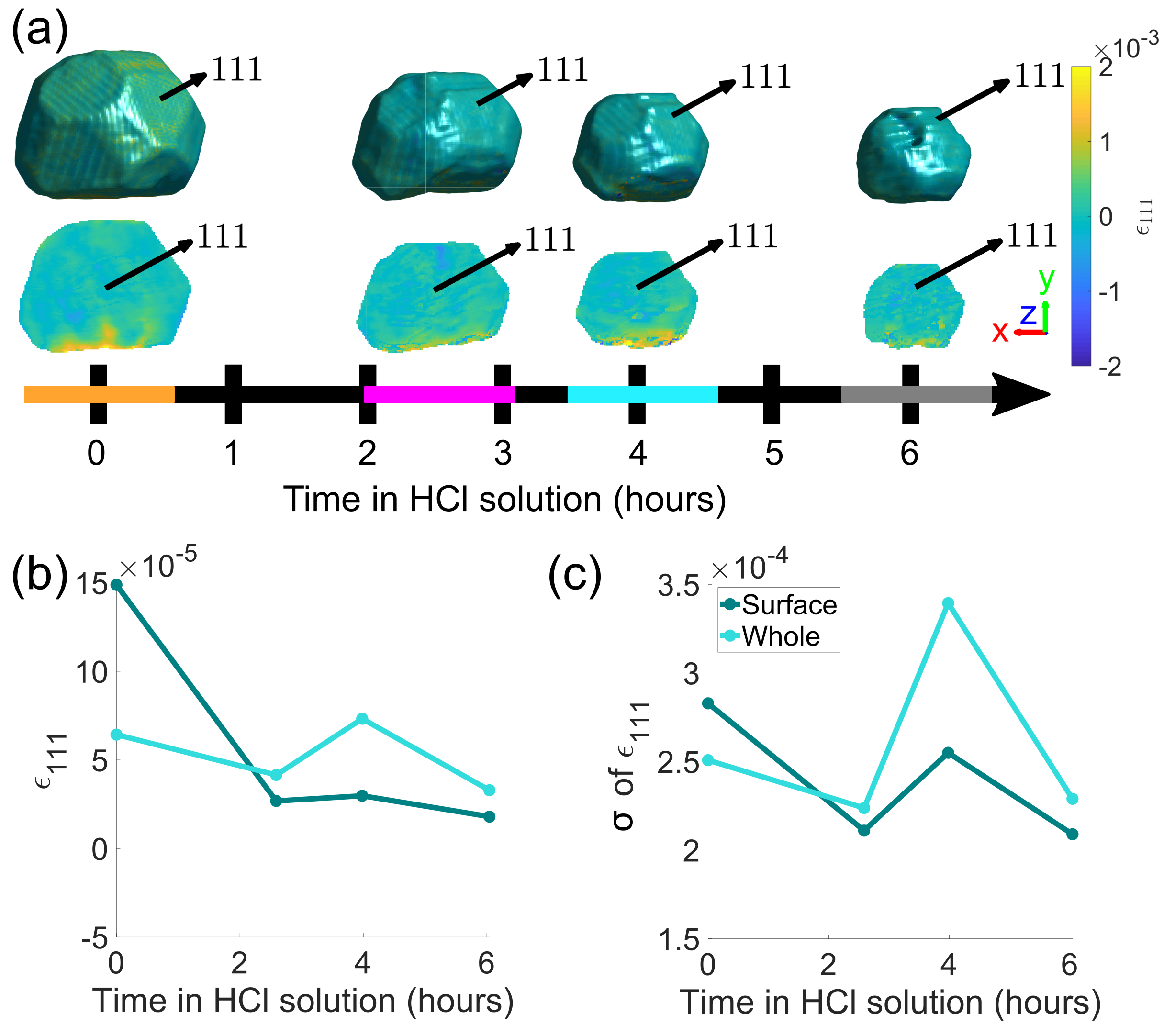}
    \caption{(a) Reconstructions of the $(111)$ reflection with an isosurface in amplitude coloured according to the local lattice strain along $[111]$ as a function of reaction time. Central slices perpendicular to the z-axis are shown in the row below. The amplitude threshold is 0.25 and the magnitude of the coordinate axes corresponds to 100 nm. The coloured bands on the time axis correspond to the reconstructions at those time points and are used throughout the article. Supplementary video SV4 rotates around the crystal during corrosion. Supplementary video SV5 shows slices through the crystal during corrosion. (b) The average surface strain compared to the average strain of the whole crystal along $[111]$. (c) The standard deviation of data points shown in (b).}
    \label{fig:Corrosion_isosurface_slice}
\end{figure}

The evolution of crystal morphology as the corrosion progresses is shown in Fig. \ref{fig:Corrosion_isosurface_slice}(a) with the displayed colour corresponding to the local lattice strain along $[111]$. Prior to corrosion, there is more tensile strain on the surface, observable in the cross sectional panels of the 3D strain tensor in Fig.\ref{fig:Tensor}. As expected, the crystal decreases in size and surface features such as the facets and edges become rougher. There is a trend where the average surface strain decreases as corrosion proceeds (Fig. \ref{fig:Corrosion_isosurface_slice} (b)), along with a subtle overall reduction in surface strain variation (Fig. \ref{fig:Corrosion_isosurface_slice} (c)). Dissolution mainly occurs at the top of the crystal, which is more exposed to the HCl solution relative to the bottom. Interestingly, in Fig. \ref{fig:Corrosion_isosurface_slice}(a), the strain at the base of the crystal does not change significantly. We attribute this to there being negligible corrosion at the base of the crystal due to the contact with the substrate and subsequent protection of this surface. Alignment of the crystal for subsequent time points was performed using the base of the crystal as a common reference point.

\begin{figure}[ht] 
    \centering
    \includegraphics[height=8.7cm]{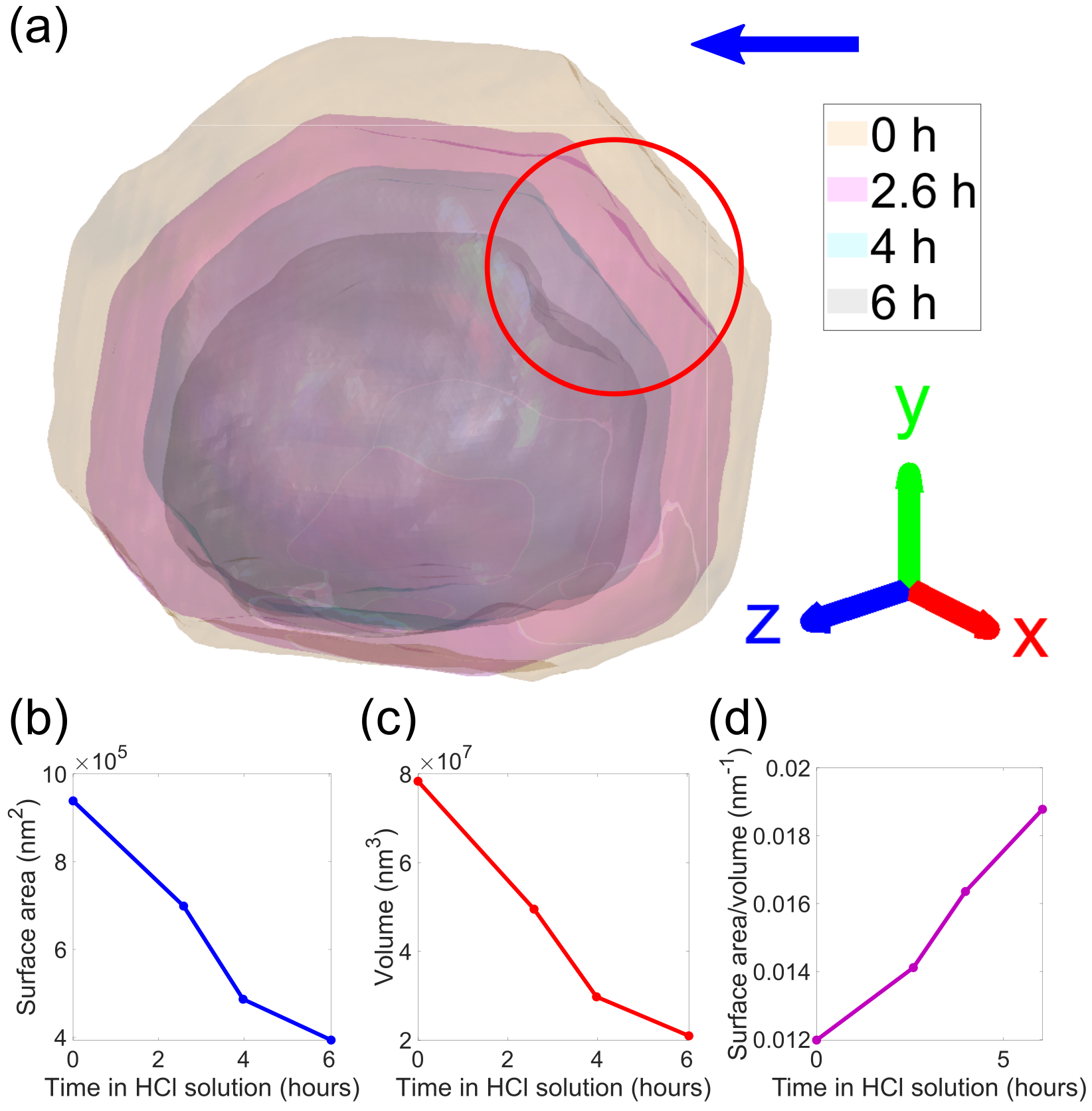}
    \caption{(a) Translucent morphologies of the $(111)$ reconstructions as a function of corrosion time. The red circle highlights the formation of a pit on the $(100)$ facet. The blue arrow shows the direction of fluid flow. The amplitude threshold is 0.25 and the magnitude of the coordinate axes corresponds to 100 nm. (b-d) The evolution of volume, surface area, and surface area to volume ratio of the crystal as a function of corrosion time, respectively.}
    \label{fig:Pit_evolution}
\end{figure}

Semi-transparent morphologies of each reconstruction during corrosion are shown superimposed in Fig. \ref{fig:Pit_evolution}. There is a pit that forms on the $(100)$ facet and enlarges during the corrosion (red circle). Pits can nucleate from highly strained morphological features or by preferential dissolution near a dislocation core--the dissolution of strains surrounding the dislocation core would reduce the overall energy of the crystal \cite{Cabrera1954,Lasaga1986}. In this case there are no noticeable heterogenities at the site of the pit and there is no noticeable dislocation at the region where the pit nucleated, shown in Fig \ref{fig:Corrosion_isosurface_slice}(a), so it is unlikely that the pit was created by crystallographic features. A hypothesis for the formation and enlargement of the pit comes from the direction of fluid flow (shown by the direction of the blue arrow): the region of the pit is most exposed to the moving liquid. However, the corrosion rate here is smaller than that reported by Horng, where Co-Fe was formed from powder metallurgy and corroded in stagnant HCl \cite{Horng1968}. 


\section{Discussion} \label{Discussion}

\subsection{Crystal morphology} 

\begin{figure}[ht] 
    \centering
    \includegraphics[height=3.8 cm]{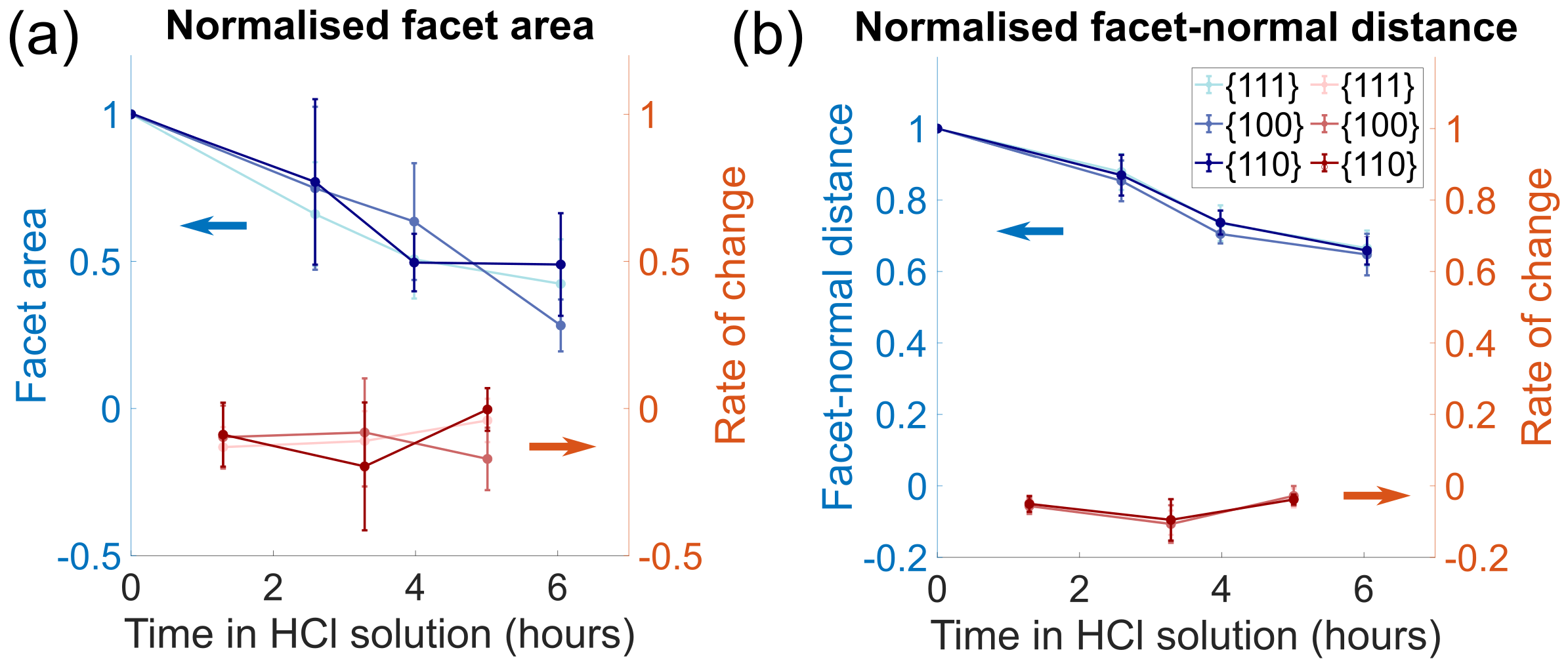}
    \caption{(a) The normalised average facet area for different families of facets (blue) and the change in normalised average facet area (red). (b) The normalised facet-normal distance (blue) from the centre of mass of the facet to the centre of mass of the crystal for each time point, and the change in normalised facet-normal distance (red). The blue and red arrows point to the y-axis for each set of lines.}
    \label{fig:Average_facet_size_normal_radii_normalised}
\end{figure}

The order of free surface energies for low index facets is $\{111\} < \{100\} < \{110\}$ based on ab initio techniques that account for the number of nearest-neighbour broken bonds \cite{Galanakis2002}. It is expected that the facets with the highest energy will dissolve most quickly. Based on the identification of facets in Appendix \ref{appendix:Facet_evolution}, the area for each facet was calculated. To characterise the change in area between different families of facets, we normalise each facet by its initial facet area and plot the change over the period of corrosion, as shown in Fig. \ref{fig:Average_facet_size_normal_radii_normalised}(a). Surprisingly, all facets sizes decrease, at a similar rate, with no clear dependence on crystallographic direction. 

To quantify the surface normal retreat for each facet family, we calculate the facet-normal distance from the centre of mass of the facet to the centre of mass of the crystal for each time point. The change in surface normal distance is shown in Fig. \ref{fig:Average_facet_size_normal_radii_normalised}(b). Consistent with the facet size change, there is negligible facet-dependent change in surface retreat.

While there is no clear difference between different families of facets when looking at their average facet sizes and average facet-normal distance to the centre of mass of the crystal, the localised dissolution rate, discussed in Section \ref{Dissolution}, provides a clearer picture of how subtle geometric constraints affect corrosion.

\subsection{Dissolution} \label{Dissolution}

The local dissolution rate was determined by computing a 3D Euclidean distance map from the morphology of the crystal, which represents the crystal-fluid interface at each time point during corrosion \cite{Noiriel2019}. Each voxel beyond the crystal surface was labelled with the smallest normal distance to the crystal surface. To determine the local dissolution at $t_i$, the 3D Euclidean distance map was computed for $t_{i+1}$ and the values were taken using the 3D coordinates of the crystal-fluid interface at $t_i$. For crystal edges and vertices, the local dissolution at $t_i$ is the distance normal to the considered edge at $t_{i+1}$. The local dissolution rate, based on the notation from Noiriel \textit{et al.} \cite{Noiriel2019}, was computed by

\begin{equation} \label{eq:corrosion}
    r'_{\mathrm{diss}} = \frac{d\mathbf{I_{fc}\cdot n}}{dt},
\end{equation}

where $\mathbf{I_{fc}}$ is the position vector and $\mathbf{n}$ is the normal to the crystal surface. 

When determining the dissolution rate, we note that the maximum change within the crystal during a scan of $\sim$10 minutes is 9 nm. This is significantly less than the average spatial resolution of the reconstructions (see Appendix \ref{appendix:phase_retrieval}). Furthermore, since the crystal remained attached to the substrate while measurements were performed, all the reconstructions were aligned such that the bottom surface of the crystal remained in the same position.

\begin{figure*}[ht] 
    \centering
    \includegraphics[height=11cm]{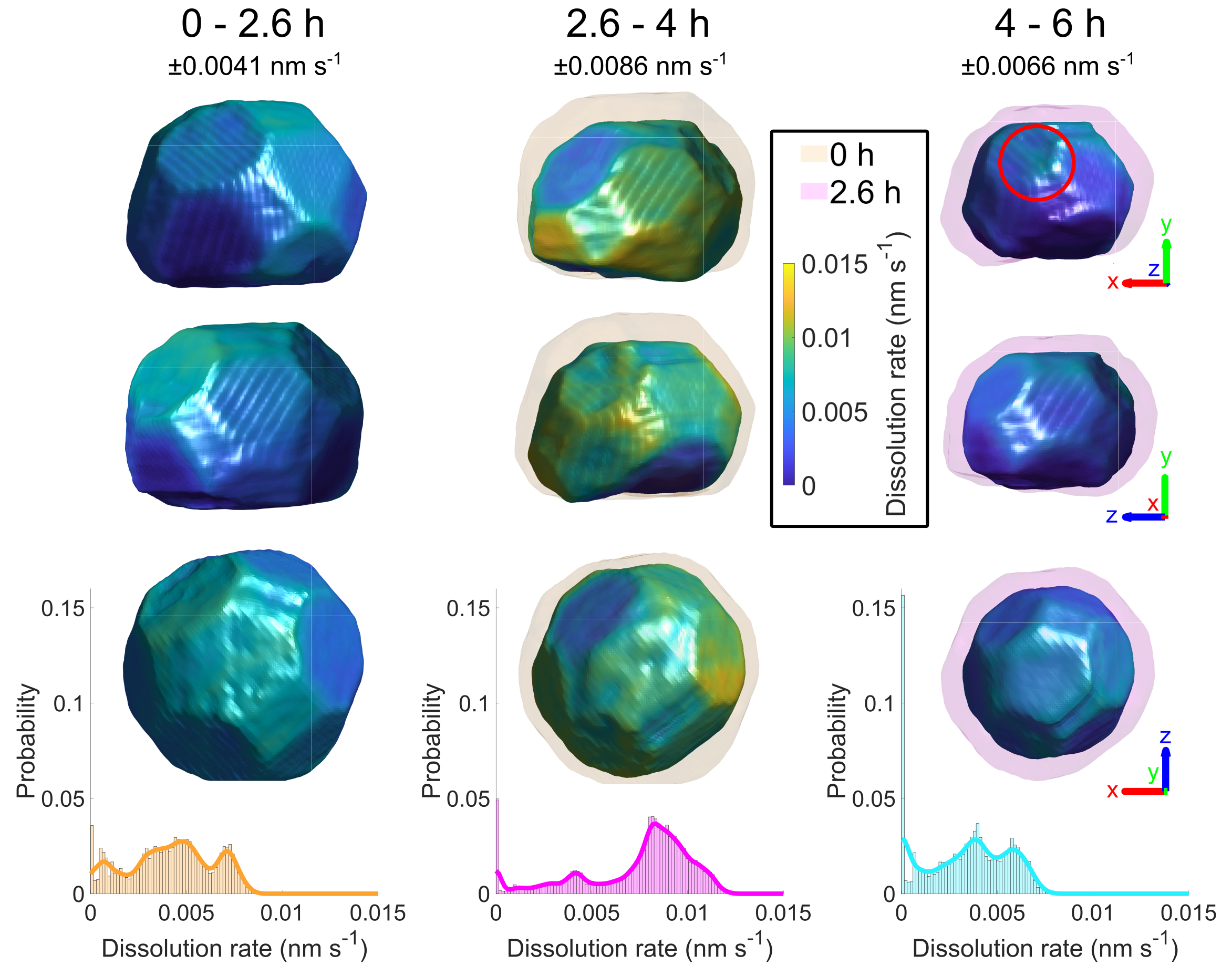}
    \caption{Crystal surface coloured according to local dissolution rate and histograms of dissolution rate between adjacent time points. The error is ± 0.0041, 0.0086, and 0.0066 nm/s for 0 - 2.6, 2.6 - 4, and 4 - 6 h respectively. The semi-transparent volumes represent the crystal morphology at the previous time step. The red circle highlights the localised increase in dissolution rate where the pit becomes noticeable at 6 h (Fig. \ref{fig:Pit_evolution}(a)). The amplitude threshold is 0.25 and the magnitude of the coordinate axes corresponds to 100 nm. Supplementary video SV6 rotates around the crystal during corrosion.}
    \label{fig:Dissolution_rate}
\end{figure*}

Fig. \ref{fig:Dissolution_rate} shows the crystal surface coloured according to the local dissolution rate. The results reveal that the dissolution of the crystal is not straightforward and evolves both spatially and temporally. From 0 - 2.6 h, the average corrosion rate is 0.0042 nm/s. From 2.6 - 4 h, the average corrosion rate approximately doubles to 0.0074 nm/s. Finally, from 4 - 6 h the corrosion rate decreases to 0.0033 nm/s. 

This low-high-low corrosion rate pattern is likely a result of the carbon coating initially present on the surface of the crystal. A previous study by Larijani \textit{et al.} shows that the corrosion of 316 stainless steel coated with 200 nm of carbon has a corrosion current density two orders of magnitude smaller and a higher corrosion potential in a prototypical fuel cell environment with a pH of 2 \cite{Larijani2011}. Thus the lower initial localised dissolution rate from 0 - 2.6 h (Fig. \ref{fig:Dissolution_rate}) is attributed to the passivation due to the carbon coating, which is slowly removed.

To understand these changes, histograms of observed dissolution rates were plotted. Note that there is a large proportion that show zero dissolution rate, corresponding to the bottom of the crystal. The first time interval from 0 - 2.6 h shows a wide variation of dissolution rates, with a broad peak at 0.0044 nm/s and a narrow peak at 0.0071 nm/s. The second time interval from 2.6 - 4 h shows one large peak at about 0.0083 nm/s and a much smaller one at 0.0041 nm/s. The final time interval from 4 - 6 h shows a double peak at about 0.0028 nm/s and 0.0057 nm/s.

A reason for the heterogeneous dissolution rate from 0 - 2.6 h is the dissolution of vertices and edges, which have been shown to be removed faster \cite{Yuan2019a} due to the large number of step and kink atoms present \cite{Luttge2013,Briese2017}. These step and kink atoms have lower coordination than facet atoms, which increases their exposure to the fluid. Furthermore, these sites can increase the reactivity by stabilising reaction intermediates and lowering energy barriers \cite{Vines2010}. 

The dissolution rate decreases again after 4 h, to similar levels as between 0 h and 2.6 h, though more homogeneously distributed. This is attributed to the crystal becoming rounder, whereby solution mass transport lowers the dissolution rate relative to that of acute corners and edges \cite{Yuan2019a}. The pit becomes significantly enlarged after 6 h (Fig. \ref{fig:Pit_evolution}). This localised higher dissolution of the pit is reflected by the rightmost peak from 4 - 6 h in Fig. \ref{fig:Dissolution_rate}.

\begin{figure}[ht] 
    \centering
    \includegraphics[height=6.5cm]{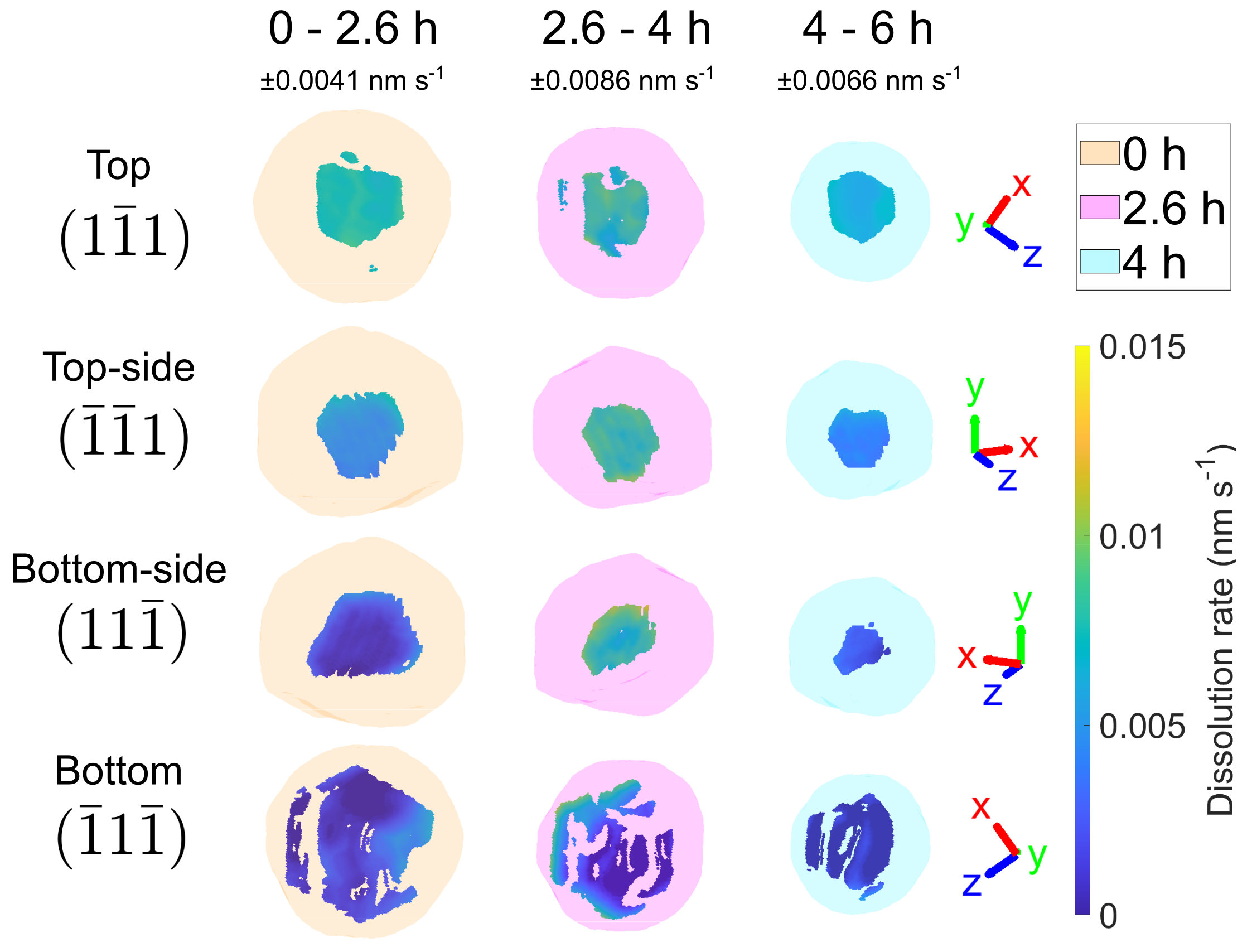}
    \caption{The dissolution of the top $(1\bar{1}1)$, upper-middle $(\bar{1}\bar{1}1)$, lower-middle $(11\bar{1})$, and bottom $(\bar{1}1\bar{1})$ facets. The error is ± 0.0041, 0.0086, and 0.0066 nm/s for 0 - 2.6, 2.6 - 4, and 4 - 6 h respectively. The semi-transparent volumes represent the crystal morphology at the time step. There is an increase in dissolution rate at 2.6 - 4h. The dissolution rate decreases the closer the facet is to the bottom of the crystal. The amplitude threshold is 0.25 and the magnitude of the coordinate axes corresponds to 100 nm.}
    \label{fig:Facet_dissolution}
\end{figure}

Fig. \ref{fig:Facet_dissolution} shows that the facet dissolution is highly position-dependent. The bottom of the crystal is protected from acid exposure by the substrate and therefore experiences the least corrosion. We observe an increase in dissolution rate from facets near the base of the crystal to the top surface, where the dissolution rate is highest. We attribute this trend to a reduction in solution mass transport near the substrate, which has also been observed in larger calcite crystals \cite{Yuan2019a}. As the crystal decreases in size, the bottom facets contribute to a larger fraction of surface area and thus biases the dissolution rate histogram towards lower values. 

\subsection{Crystal strain} 

\begin{figure}[ht] 
    \centering
    \includegraphics[height=4.8cm]{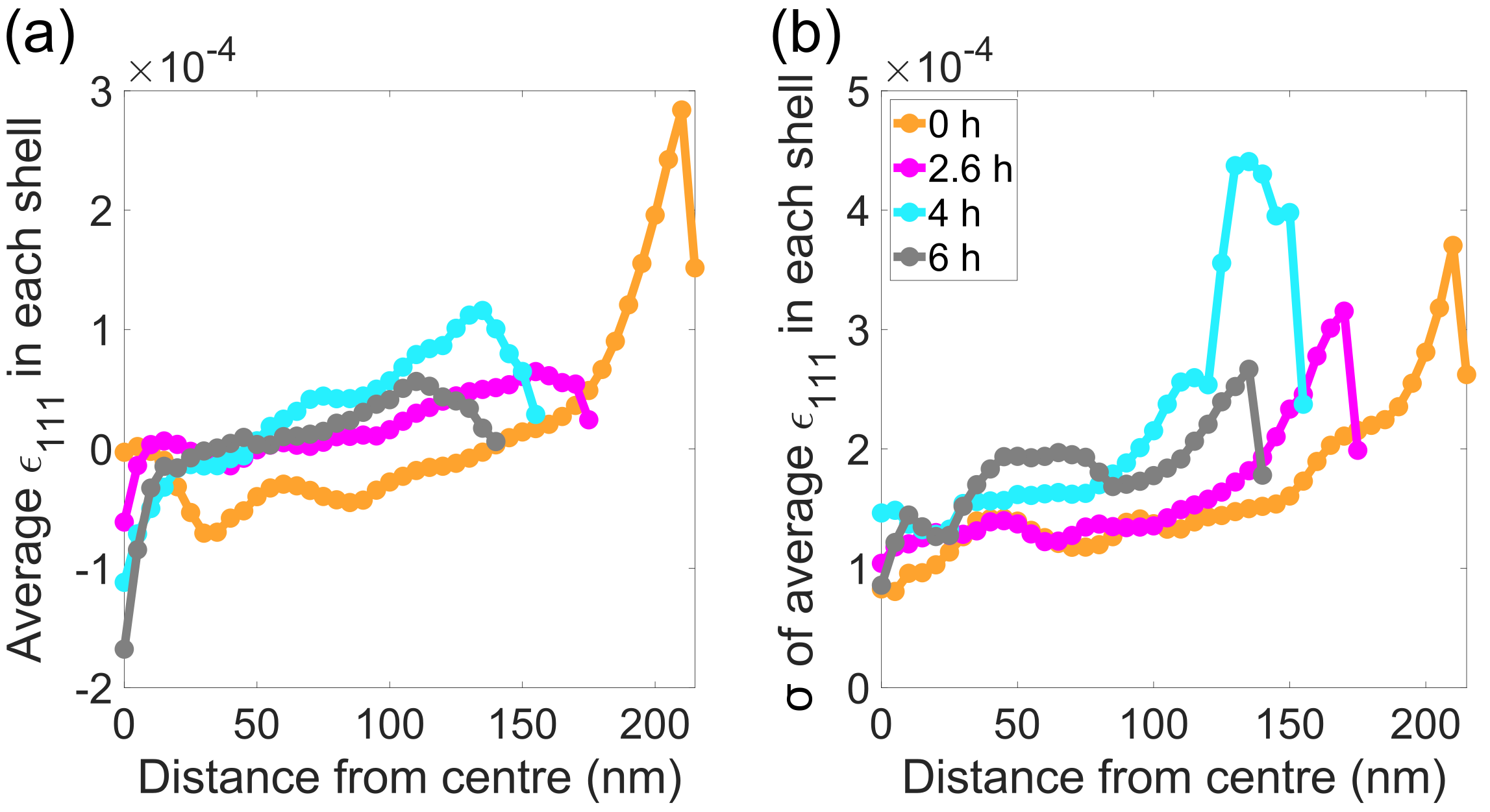}
    \caption{(a) The average strain along $[111]$ as a function of distance from centre of mass of the crystal at each time point. (b) The standard deviation of the average strain in (a).}
    \label{fig:Average_strain_shell}
\end{figure}

We hypothosize that the heterogeneous dissolution rate from 0 - 2.6 h is due to the higher surface strain, which progressively decreases as the reaction proceeds (Fig. \ref{fig:Corrosion_isosurface_slice}(b)). Regions of higher strain have been reported to increase the dissolution rate due to increased internal energy \cite{Morel2001,Schott1989,Gutman1994}. The average strain inside the crystal presented in Fig. \ref{fig:Corrosion_isosurface_slice}(b and c) is analysed in more detail in Fig \ref{fig:Average_strain_shell}, which shows the average strain along $[111]$ for concentric shells within the crystal. Each shell is one voxel thick, centred about the crystal's centre of mass, and is scaled down from the surface morphology. The surface strain is clearly highest before corrosion, and decreases as the reaction progresses. With increasing corrosion time, the average strain increases within the volume of the crystal (Fig. \ref{fig:Average_strain_shell}(a)) and the strain distributions inside the crystal becomes less homogeneous (Fig. \ref{fig:Average_strain_shell}(b)). 

After the edges, vertices, and most strained regions of the crystal dissolve in the first 4 hours, the dissolution rate decreases. This could be due to the core being more stable after more highly strained outer layers of the crystal have been removed, which agrees with the observations of Clark \textit{et al.} \cite{Clark2015}. Interestingly, the standard deviation of the strain in each shell becomes smaller closer to the centre (Fig. \ref{fig:Average_strain_shell}(b)). However, the standard deviation at a given distance from the centre increases with advancing corrosion. This suggests that the strain within the crystal becomes more heterogeneous as the crystal dissolves.

Based on post-corrosion EDX analysis, there is no significant dealloying of Fe, the less noble element, from the present crystals. However, a change in Cr and Ni impurities after corrosion suggests that there may have been preferential leaching (see Appendix \ref{appendix:EDX_results}). 

Here, the BCDI rocking scans required just under 10 minutes to collect. For \textit{ex situ} measurements, repeated scans are typically aligned and summed to improve reconstruction quality. However, due to the constant evolution of the crystal, only slow changes in corrosion can be probed. Pre-corrosion measurements showed no evidence of beam induced sample damage based on the high cross-correlation of repeated scans. We therefore attribute subsequent changes of the sample during \textit{in situ} corrosion entirely to the corrosion of the sample itself. X-ray radiation can cause excess heat and the ionisation of water \cite{Mesu2006}, both of which can accelerate the corrosion rate. This must be considered when using \textit{in situ} BCDI to study wet chemistry processes.

\section{Conclusions}
This prototypical study demonstrates the feasibility of using BCDI to understand how strain and morphology contribute to nanoscale corrosion rate. We note that analysing localised dissolution is critical to understanding how defects effect the nature of corrosion. Here, we observe that localised corrosion rates can be up to two times higher than the average corrosion rate of the crystal. Regions such as vertices and edges are shown to corrode faster than facets, leading to a progressively rounder crystal morphology. Furthermore, corrosion leads to a relaxation of surface strain. These findings would not be apparent when analysing changes in average facet size and average facet-normal distance. Finally, we demonstrate that corrosion at the sub-micron level appears to be heavily dependent on environmental conditions, as the top facets that are more exposed to the flowing fluid corrode more quickly than bottom facets.

BCDI uniquely allows the \textit{in situ} study of how corrosion progresses and interacts with the crystal morphology, defects and strain distribution. This insight to transient structures is not accessible via any other technique and can be crucial in understanding the driving mechanisms of corrosion. In turn, this may enable the design of better corrosion-resistant materials in the future, e.g. by enabling the selection of corrosion-resistant textures in polycrystalline samples.

\section{Data availability}
The processed diffraction patterns, final reconstructions, and data analysis scripts are publicly available at 10.5281/zenodo.5548084.

\appendix

\section{Phase retrieval} \label{appendix:phase_retrieval}
The reconstruction process of CXDPs was executed in stages, using the output from the previous stage to seed the next phasing stage (listed below). Initially, CXDPs were padded to a size of $256 \times 256 \times 144$ voxels.
\begin{enumerate}
    \item Each reconstruction was seeded with a random guess. A guided phasing approach \cite{Chen2007} with 40 individuals and four generations was used with a geometric average breeding mode. For each generation and population, a block of 20 error reduction (ER) and 180 hybrid input-output (HIO) iterations, with $\beta = 0.9$, was repeated three times. This was followed by 20 ER iterations to return the final object. The shrinkwrap algorithm \cite{Marchesini2003} with a threshold of 0.1 was used to update the real-space support every iteration. The $\sigma$ for the Gaussian kernel shrinkwrap smoothing for each generation was $\sigma = 2.0, 1.5, 1.0, 1.0$ respectively. The best reconstruction was determined using a sharpness criterion, as it is the most appropriate metric for crystals containing defects \cite{Ulvestad2017b}. The two worst reconstructions were removed after each generation.
    
    \item The reconstruction was seeded with the output from stage 1. This stage was identical to stage 1, except now we account for the longitudinal and transverse partial coherence using a mutual coherence function (MCF) \cite{Clark2012}. Starting from the 50th iteration, the MCF was updated every 50 iterations using 20 iterations of the Richardson–Lucy deconvolution algorithm\cite{Richardson1972}.
    
    \item The reconstruction was seeded with the output from stage 2. This stage was identical to stage 2, except now we use a guided phasing approach with 10 individuals and four generations. For each generation and population, a block of 20 error reduction (ER) and 180 hybrid input-output (HIO) iterations, with $\beta = 0.9$, was repeated 15 times, followed by 1000 ER iterations. The final 50 iterates were averaged to produce the final image.
\end{enumerate}

The average 3D spatial resolution was 30 nm. This was determined by differentiating the electron density amplitude across the crystal/acid interface for the five reflection directions and fitting a Gaussian to each of the profiles. The reported spatial resolution is the averaged full width at half maximum of the Gaussian profiles. The average 3D spatial resolution for each reconstruction during corrosion was 23, 31, 30 and 38 nm at 0, 2.6, 4 and 6 h respectively.

\section{Strain calculations} \label{appendix:strain_calculations}
We accounted for arbitrary phase offsets by establishing a cubic volume (with a size of half of the cubed root of the crystal volume) centred about the centre of mass as the zero-phase reference for every reconstruction. The resulting reconstructions were transformed from detector conjugated space to orthogonal sample space with a voxel size of $5 \times 5 \times 5 \mathrm{\ nm^3}$. 


Phase ramps were eliminated by centring the Fourier transform of the complex electron density in detector conjugated space and sample space. The strain, $\epsilon\mathbf{(r)}$, projected along the $hkl$ Bragg vector was found by differentiation,

\begin{equation} \label{eq:strain}
    \epsilon_{hkl}\mathbf{(r)} = \nabla\mathbf{\psi_\mathit{hkl}(r)}\cdot\frac{\mathbf{Q_\mathit{hkl}}}{|\mathbf{Q_\mathit{hkl}}|^2}.
\end{equation}


The full 3D lattice strain tensor, $\mathbf{\varepsilon}(\mathbf{r})$, and rotation tensor, $\mathbf{\omega}(\mathbf{r})$, are obtained by \cite{Constantinescu2008}

\begin{equation} \label{eq:lattice_strain_tensor}
    \mathbf{\varepsilon}(\mathbf{r}) = \frac{1}{2}\left\{\nabla \mathbf{u}(\mathbf{r})+[\nabla\mathbf{u}(\mathbf{r})]^\mathrm{T}\right\} \mathrm{, and}
\end{equation}

\begin{equation} \label{eq:rotation_tensor}
    \mathbf{\omega}(\mathbf{r}) = \frac{1}{2}\left\{\nabla \mathbf{u}(\mathbf{r})-[\nabla\mathbf{u}(\mathbf{r})]^\mathrm{T}\right\},
\end{equation}

which rely on the reconstruction of $\mathbf{u}(\mathbf{r})$. With a single BCDI measurement, we can determine one component of the displacement field. For multi-reflection BDCI (MBCDI), if at least three linearly independent reflections are measured, $\mathbf{u}(\mathbf{r})$ can be determined by minimising the least-squares error \cite{Hofmann2017b,Newton2010},

\begin{equation} \label{eq:least-squares_displacement}
    E(\mathbf{r}) = \sum_{hkl}[\mathbf{Q}_{hkl}\cdot\mathbf{u}(\mathbf{r})-\psi_{hkl}(\mathbf{r})]^2,
\end{equation}

for every voxel in the sample. Here, we use the modified a approach by Hofmann \textit{et al.} \cite{Hofmann2020}, in which case, the squared error between phase gradients is minimised:

\begin{equation} \label{eq:least-squares_gradient}
    E(\mathbf{r})_j = \sum_{hkl,j} \left[\mathbf{Q}_{hkl}\cdot\frac{\partial \mathbf{u}(\mathbf{r})}{\partial j}-\frac{\partial \psi_{hkl}(\mathbf{r})}{\partial i}\right]^2,
\end{equation}

where $j$ corresponds to the spatial $x$, $y$, or $z$ coordinate, to find $\nabla \mathbf{u}(\mathbf{r})$ directly for the computation of $\mathbf{\varepsilon}(\mathbf{r})$, and $\mathbf{\omega}(\mathbf{r})$ in Eqs. \ref{eq:lattice_strain_tensor} and \ref{eq:rotation_tensor} respectively.

\section{Facet evolution} \label{appendix:Facet_evolution}

\begin{figure}[ht] 
    \centering
    \includegraphics[height=6.5cm]{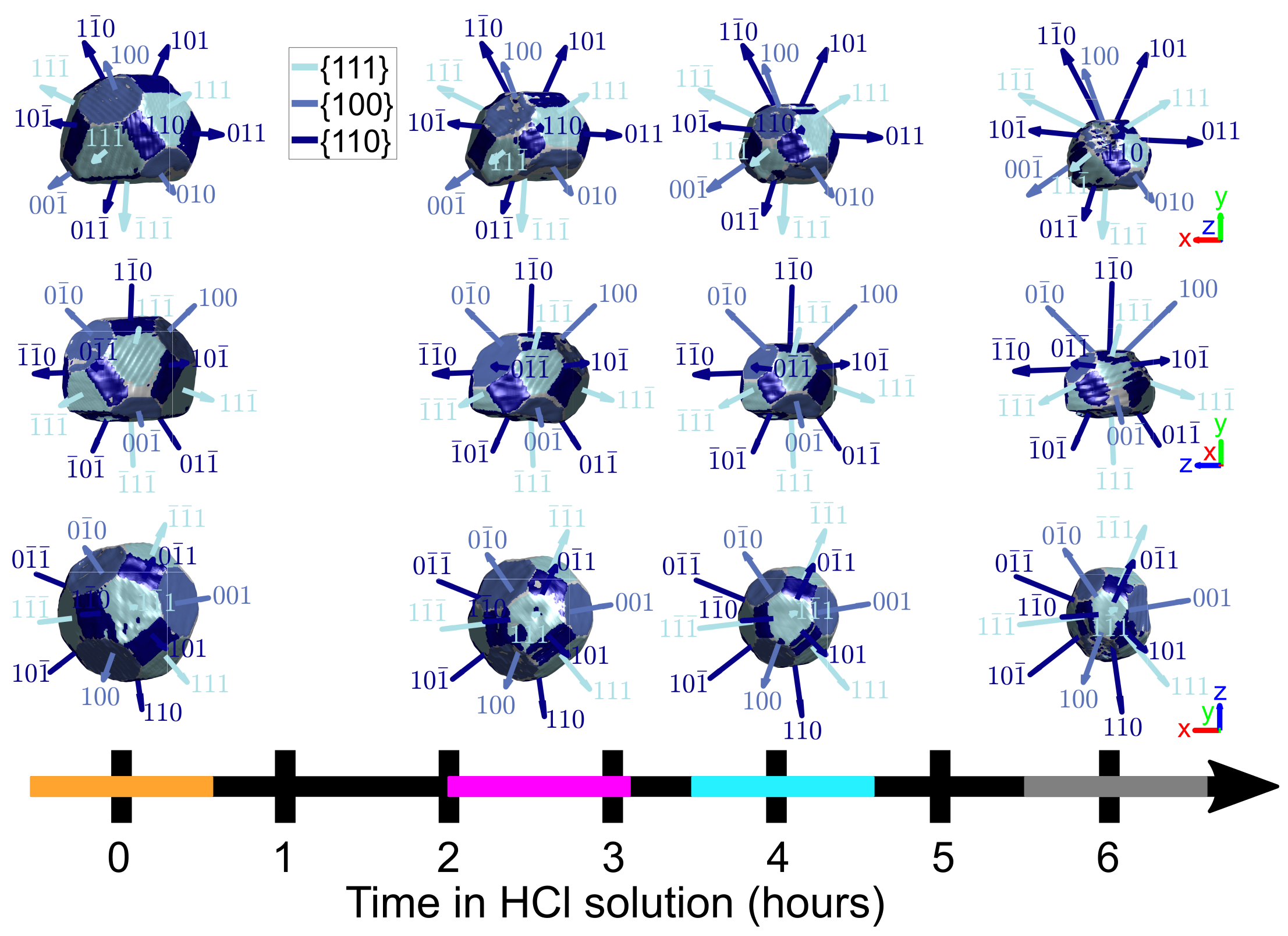}
    \caption{The identification of facets on the crystal surface as a function of corrosion time. Shown are the x-y, x-z, and y-z planes. The coloured bands on the time axis correspond to the reconstructions at those time points and are used throughout the article. For the crystals, the amplitude threshold is 0.25 and the magnitude of the coordinate axes corresponds to 100 nm. Supplementary video SV7 rotates around the crystal with facets identified at each time point during corrosion.}
    \label{fig:Facet_evolution}
\end{figure}

Fig. \ref{fig:Facet_evolution} highlights the evolution of facets. Facet areas are identified by selecting the surface elements that have an isonormal within a $19 \degree$ angular threshold of a given facet normal unit vector. The facet normal vectors were calculated from the orientation matrix obtained by minimising the error between the five known reflections in sample space. In principle, a maximum angular threshold of $27.4 \degree$ could be used, which is half the angle between $\{111\}$ and $\{100\}$ planes. Our chosen tolerance is below this maximum since not all facets are fully developed, and the edges and vertices of the crystal appear rounded and thus should not be classified as part of a facet. The facet areas were calculated by summing the areas of all surface elements that falls within the angular threshold for facet identification.

\section{EDX results} \label{appendix:EDX_results}

\begin{figure}[ht] 
    \centering
    \includegraphics[height=8cm]{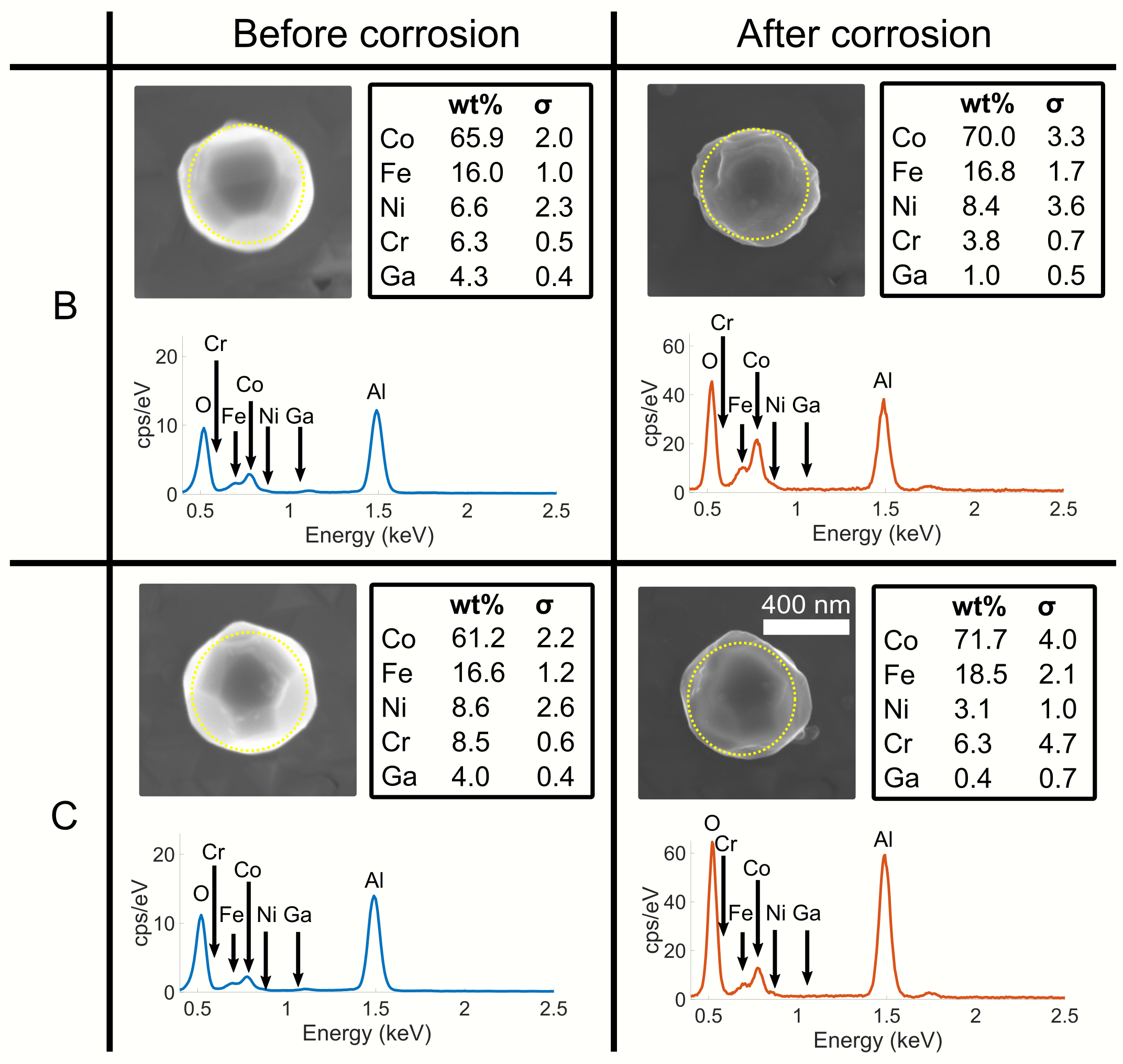}
    \caption{A comparison between two other crystals (B and C) on the substrate before and after corrosion using SEM and EDX. The EDX spectra before corrosion (blue) were collected on a ZEISS EVO with an acceleration voltage of 10 kV, consistent with the spectrum shown in Fig. \ref{fig:SEM_pic}(d). EDX spectra after corrosion (orange) were collected on a ZEISS Merlin analytical SEM with an acceleration voltage of 10 kV. The dotted yellow line represents the region measured using EDX.}
    \label{fig:EDX_comparison}
\end{figure}

Fig. \ref{fig:EDX_comparison} compares two other crystals (B and C) on the substrate before and after corrosion using SEM images and EDX spectra. There is no obvious indication of dealloying, as the ratio of Co and Fe does not show significant change. Preferential leaching of Cr or Ni may occur, but is not considered further since the concentrations within the samples are low. The concentration of Ga decreased, suggesting the dissolution of most of the FIB-contaminated parts of the crystal.

\begin{figure}[ht] 
    \centering
    \includegraphics[height=2.7cm]{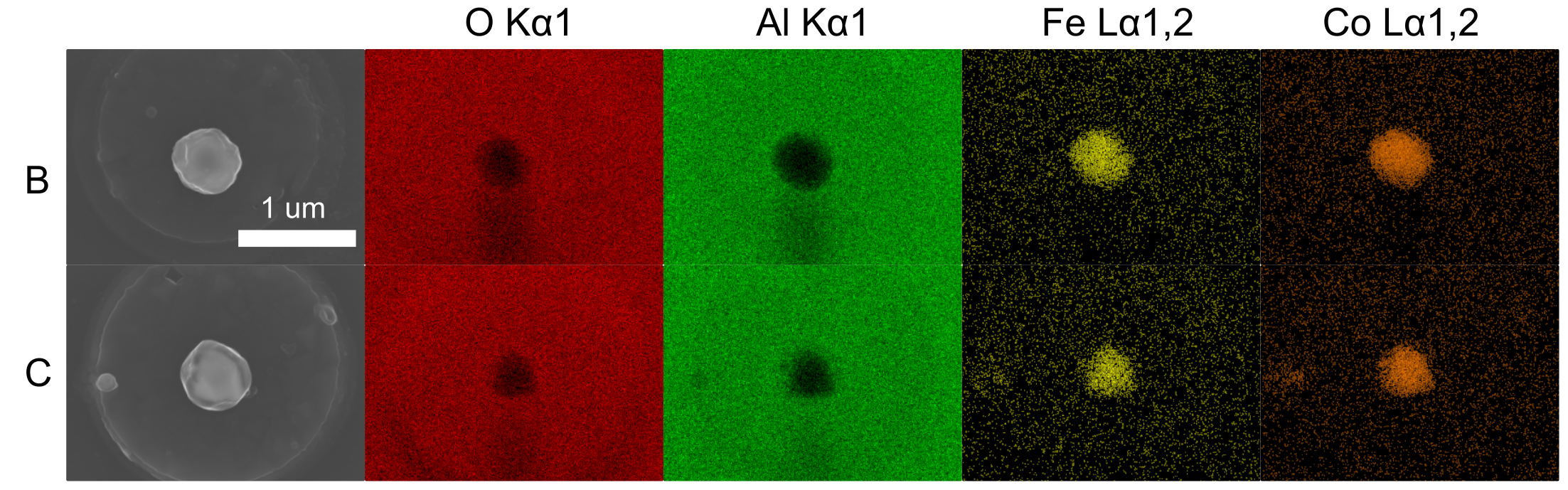}
    \caption{EDX elemental analysis maps of the two crystals (B and C) shown in Fig. \ref{fig:EDX_comparison}. Shown are 2D signals for the substrate, Al\textsubscript{2}O\textsubscript{3}, and primary Co-Fe emission lines. The maps were collected using a ZEISS Merlin analytical SEM equipped with an in-lens detector with an acceleration voltage of 10 kV.}
    \label{fig:EDX_map}
\end{figure}

Fig. \ref{fig:EDX_map} shows homogeneous elemental distribution of the two crystals shown in Fig. \ref{fig:EDX_comparison} after corrosion. 

\section{Supplementary video descriptions} \label{appendix:supplementary_videos}
A brief description of each supplementary video is listed below.
\begin{itemize}
\item SV1 shows x-y plane slices through the lattice strain and rotation tensors in Fig. \ref{fig:Tensor}.
\item SV2 shows y-z plane slices through the lattice strain and rotation tensors in Fig. \ref{fig:Tensor}.
\item SV3 shows z-x plane slices through the lattice strain and rotation tensors in Fig. \ref{fig:Tensor}.
\item SV4 shows the rotation of the crystal coloured by strain at various time points during corrosion, based on Fig. \ref{fig:Corrosion_isosurface_slice}(a).
\item SV5 shows x-y plane slices through the crystal at various time points during corrosion, based on Fig. \ref{fig:Corrosion_isosurface_slice}(a).
\item SV6 shows the rotation of the crystal coloured by strain at various time points during corrosion, based on Fig. \ref{fig:Dissolution_rate}.
\item SV7 shows the rotation of the crystal coloured by facets at various time points during corrosion, based on Fig. \ref{fig:Facet_evolution}.
\end{itemize}
\

\section*{Conflicts of interest}
There are no conflicts to declare.

\section*{Acknowledgements}
D.Y, N.W.P and F.H. acknowledge funding from the European Research Council under the European Union's Horizon 2020 research and innovation programme (grant agreement No 714697 and under the Marie Skłodowska-Curie Actions grant agreement No. 884104 PSI-FELLOW-III-3i). K.S. acknowledges funding from the General Sir John Monash Foundation. The authors acknowledge use of characterisation facilities at the David Cockayne Centre for Electron Microscopy, Department of Materials, University of Oxford and use of the Advanced Research Computing (ARC) facility at the University of Oxford \cite{Richards2015}. X-ray diffraction experiments were performed at the Advanced Photon Source, a US Department of Energy (DOE) Office of Science User Facility operated for the DOE Office of Science by Argonne National Laboratory under Contract No. DE-AC02-06CH11357.



\balance


\bibliography{rsc} 

\providecommand*{\mcitethebibliography}{\thebibliography}
\csname @ifundefined\endcsname{endmcitethebibliography}
{\let\endmcitethebibliography\endthebibliography}{}
\begin{mcitethebibliography}{51}
\providecommand*{\natexlab}[1]{#1}
\providecommand*{\mciteSetBstSublistMode}[1]{}
\providecommand*{\mciteSetBstMaxWidthForm}[2]{}
\providecommand*{\mciteBstWouldAddEndPuncttrue}
  {\def\EndOfBibitem{\unskip.}}
\providecommand*{\mciteBstWouldAddEndPunctfalse}
  {\let\EndOfBibitem\relax}
\providecommand*{\mciteSetBstMidEndSepPunct}[3]{}
\providecommand*{\mciteSetBstSublistLabelBeginEnd}[3]{}
\providecommand*{\EndOfBibitem}{}
\mciteSetBstSublistMode{f}
\mciteSetBstMaxWidthForm{subitem}
{(\emph{\alph{mcitesubitemcount}})}
\mciteSetBstSublistLabelBeginEnd{\mcitemaxwidthsubitemform\space}
{\relax}{\relax}

\bibitem[Lasaga(1998)]{Lasaga1998}
A.~C. Lasaga, \emph{{Kinetic theory in the earth sciences}}, Princeton
  University Press, Princeton, N.J., 1998\relax
\mciteBstWouldAddEndPuncttrue
\mciteSetBstMidEndSepPunct{\mcitedefaultmidpunct}
{\mcitedefaultendpunct}{\mcitedefaultseppunct}\relax
\EndOfBibitem
\bibitem[Şahin and {\"{U}}beyli(2008)]{Sahin2008}
S.~Şahin and M.~{\"{U}}beyli, \emph{Journal of Fusion Energy}, 2008,
  \textbf{27}, 271--277\relax
\mciteBstWouldAddEndPuncttrue
\mciteSetBstMidEndSepPunct{\mcitedefaultmidpunct}
{\mcitedefaultendpunct}{\mcitedefaultseppunct}\relax
\EndOfBibitem
\bibitem[Strasser \emph{et~al.}(2010)Strasser, Koh, Anniyev, Greeley, More, Yu,
  Liu, Kaya, Nordlund, Ogasawara, Toney, and Nilsson]{Strasser2010}
P.~Strasser, S.~Koh, T.~Anniyev, J.~Greeley, K.~More, C.~Yu, Z.~Liu, S.~Kaya,
  D.~Nordlund, H.~Ogasawara, M.~F. Toney and A.~Nilsson, \emph{Nature
  Chemistry}, 2010, \textbf{2}, 454--460\relax
\mciteBstWouldAddEndPuncttrue
\mciteSetBstMidEndSepPunct{\mcitedefaultmidpunct}
{\mcitedefaultendpunct}{\mcitedefaultseppunct}\relax
\EndOfBibitem
\bibitem[Fischer \emph{et~al.}(2012)Fischer, Arvidson, and
  L{\"{u}}ttge]{Fischer2012}
C.~Fischer, R.~S. Arvidson and A.~L{\"{u}}ttge, \emph{Geochimica et
  Cosmochimica Acta}, 2012, \textbf{98}, 177--185\relax
\mciteBstWouldAddEndPuncttrue
\mciteSetBstMidEndSepPunct{\mcitedefaultmidpunct}
{\mcitedefaultendpunct}{\mcitedefaultseppunct}\relax
\EndOfBibitem
\bibitem[Fischer \emph{et~al.}(2014)Fischer, Kurganskaya, Sch{\"{a}}fer, and
  L{\"{u}}ttge]{Fischer2014}
C.~Fischer, I.~Kurganskaya, T.~Sch{\"{a}}fer and A.~L{\"{u}}ttge, \emph{Applied
  Geochemistry}, 2014, \textbf{43}, 132--157\relax
\mciteBstWouldAddEndPuncttrue
\mciteSetBstMidEndSepPunct{\mcitedefaultmidpunct}
{\mcitedefaultendpunct}{\mcitedefaultseppunct}\relax
\EndOfBibitem
\bibitem[Noiriel \emph{et~al.}(2020)Noiriel, Oursin, and Daval]{Noiriel2020}
C.~Noiriel, M.~Oursin and D.~Daval, \emph{Geochimica et Cosmochimica Acta},
  2020, \textbf{273}, 1--25\relax
\mciteBstWouldAddEndPuncttrue
\mciteSetBstMidEndSepPunct{\mcitedefaultmidpunct}
{\mcitedefaultendpunct}{\mcitedefaultseppunct}\relax
\EndOfBibitem
\bibitem[Saldi \emph{et~al.}(2017)Saldi, Voltolini, and Knauss]{Saldi2017}
G.~D. Saldi, M.~Voltolini and K.~G. Knauss, \emph{Geochimica et Cosmochimica
  Acta}, 2017, \textbf{206}, 94--111\relax
\mciteBstWouldAddEndPuncttrue
\mciteSetBstMidEndSepPunct{\mcitedefaultmidpunct}
{\mcitedefaultendpunct}{\mcitedefaultseppunct}\relax
\EndOfBibitem
\bibitem[Noiriel \emph{et~al.}(2019)Noiriel, Oursin, Saldi, and
  Haberth{\"{u}}r]{Noiriel2019}
C.~Noiriel, M.~Oursin, G.~Saldi and D.~Haberth{\"{u}}r, \emph{ACS Earth and
  Space Chemistry}, 2019, \textbf{3}, 100--108\relax
\mciteBstWouldAddEndPuncttrue
\mciteSetBstMidEndSepPunct{\mcitedefaultmidpunct}
{\mcitedefaultendpunct}{\mcitedefaultseppunct}\relax
\EndOfBibitem
\bibitem[Cabrera \emph{et~al.}(1954)Cabrera, Levine, and Plaskett]{Cabrera1954}
N.~Cabrera, M.~M. Levine and J.~S. Plaskett, \emph{Phys. Rev.}, 1954,
  \textbf{96}, 1153\relax
\mciteBstWouldAddEndPuncttrue
\mciteSetBstMidEndSepPunct{\mcitedefaultmidpunct}
{\mcitedefaultendpunct}{\mcitedefaultseppunct}\relax
\EndOfBibitem
\bibitem[Lasaga and Blum(1986)]{Lasaga1986}
A.~C. Lasaga and A.~E. Blum, \emph{Geochimica et Cosmochimica Acta}, 1986,
  \textbf{50}, 2363--2379\relax
\mciteBstWouldAddEndPuncttrue
\mciteSetBstMidEndSepPunct{\mcitedefaultmidpunct}
{\mcitedefaultendpunct}{\mcitedefaultseppunct}\relax
\EndOfBibitem
\bibitem[Fischer and L{\"{u}}ttge(2007)]{Fischer2007}
C.~Fischer and A.~L{\"{u}}ttge, \emph{American Journal of Science}, 2007,
  \textbf{307}, 955--973\relax
\mciteBstWouldAddEndPuncttrue
\mciteSetBstMidEndSepPunct{\mcitedefaultmidpunct}
{\mcitedefaultendpunct}{\mcitedefaultseppunct}\relax
\EndOfBibitem
\bibitem[Hillner \emph{et~al.}(1992)Hillner, Manne, Gratz, and
  Hansma]{Hillner1992}
P.~E. Hillner, S.~Manne, A.~J. Gratz and P.~K. Hansma, \emph{Ultramicroscopy},
  1992, \textbf{42-44}, 1387--1393\relax
\mciteBstWouldAddEndPuncttrue
\mciteSetBstMidEndSepPunct{\mcitedefaultmidpunct}
{\mcitedefaultendpunct}{\mcitedefaultseppunct}\relax
\EndOfBibitem
\bibitem[Brand \emph{et~al.}(2017)Brand, Feng, and Bullard]{Brand2017}
A.~S. Brand, P.~Feng and J.~W. Bullard, \emph{Geochimica et Cosmochimica Acta},
  2017, \textbf{213}, 317--329\relax
\mciteBstWouldAddEndPuncttrue
\mciteSetBstMidEndSepPunct{\mcitedefaultmidpunct}
{\mcitedefaultendpunct}{\mcitedefaultseppunct}\relax
\EndOfBibitem
\bibitem[Ueta \emph{et~al.}(2013)Ueta, Satoh, Nishimura, Ueda, and
  Tsukamoto]{Ueta2013}
S.~Ueta, H.~Satoh, Y.~Nishimura, A.~Ueda and K.~Tsukamoto, \emph{Journal of
  Crystal Growth}, 2013, \textbf{363}, 294--299\relax
\mciteBstWouldAddEndPuncttrue
\mciteSetBstMidEndSepPunct{\mcitedefaultmidpunct}
{\mcitedefaultendpunct}{\mcitedefaultseppunct}\relax
\EndOfBibitem
\bibitem[Chee \emph{et~al.}(2015)Chee, Pratt, Hattar, Duquette, Ross, and
  Hull]{Chee2015}
S.~W. Chee, S.~H. Pratt, K.~Hattar, D.~Duquette, F.~M. Ross and R.~Hull,
  \emph{Chemical Communications}, 2015, \textbf{51}, 168--171\relax
\mciteBstWouldAddEndPuncttrue
\mciteSetBstMidEndSepPunct{\mcitedefaultmidpunct}
{\mcitedefaultendpunct}{\mcitedefaultseppunct}\relax
\EndOfBibitem
\bibitem[Grogan \emph{et~al.}(2014)Grogan, Schneider, Ross, and
  Bau]{Grogan2013}
J.~M. Grogan, N.~M. Schneider, F.~M. Ross and H.~H. Bau, \emph{Nano Letters},
  2014, \textbf{14}, 359--364\relax
\mciteBstWouldAddEndPuncttrue
\mciteSetBstMidEndSepPunct{\mcitedefaultmidpunct}
{\mcitedefaultendpunct}{\mcitedefaultseppunct}\relax
\EndOfBibitem
\bibitem[Chen \emph{et~al.}(2014)Chen, Reischl, Spijker, Holmberg, Laasonen,
  and Foster]{Chen2014}
J.~C. Chen, B.~Reischl, P.~Spijker, N.~Holmberg, K.~Laasonen and A.~S. Foster,
  \emph{Physical Chemistry Chemical Physics}, 2014, \textbf{16},
  22545--22554\relax
\mciteBstWouldAddEndPuncttrue
\mciteSetBstMidEndSepPunct{\mcitedefaultmidpunct}
{\mcitedefaultendpunct}{\mcitedefaultseppunct}\relax
\EndOfBibitem
\bibitem[Yuan \emph{et~al.}(2019)Yuan, Starchenko, Lee, {De Andrade}, Gursoy,
  Sturchio, and Fenter]{Yuan2019a}
K.~Yuan, V.~Starchenko, S.~S. Lee, V.~{De Andrade}, D.~Gursoy, N.~C. Sturchio
  and P.~Fenter, \emph{ACS Earth and Space Chemistry}, 2019, \textbf{3},
  833--843\relax
\mciteBstWouldAddEndPuncttrue
\mciteSetBstMidEndSepPunct{\mcitedefaultmidpunct}
{\mcitedefaultendpunct}{\mcitedefaultseppunct}\relax
\EndOfBibitem
\bibitem[Schott \emph{et~al.}(1989)Schott, Brantley, Crerar, Guy, Borcsik, and
  Willaime]{Schott1989}
J.~Schott, S.~Brantley, D.~Crerar, C.~Guy, M.~Borcsik and C.~Willaime,
  \emph{Geochimica et Cosmochimica Acta}, 1989, \textbf{53}, 373--382\relax
\mciteBstWouldAddEndPuncttrue
\mciteSetBstMidEndSepPunct{\mcitedefaultmidpunct}
{\mcitedefaultendpunct}{\mcitedefaultseppunct}\relax
\EndOfBibitem
\bibitem[Vogel and Lovell(1956)]{Vogel1956}
F.~L. Vogel and L.~C. Lovell, \emph{Journal of Applied Physics}, 1956,
  \textbf{27}, 1413--1415\relax
\mciteBstWouldAddEndPuncttrue
\mciteSetBstMidEndSepPunct{\mcitedefaultmidpunct}
{\mcitedefaultendpunct}{\mcitedefaultseppunct}\relax
\EndOfBibitem
\bibitem[Gutman(1994)]{Gutman1994}
E.~M. Gutman, \emph{{Mechanochemistry of Solid Surfaces}}, WORLD SCIENTIFIC,
  1994, p. 332\relax
\mciteBstWouldAddEndPuncttrue
\mciteSetBstMidEndSepPunct{\mcitedefaultmidpunct}
{\mcitedefaultendpunct}{\mcitedefaultseppunct}\relax
\EndOfBibitem
\bibitem[Morel and {Den Brok}(2001)]{Morel2001}
J.~Morel and S.~W. {Den Brok}, \emph{Journal of Crystal Growth}, 2001,
  \textbf{222}, 637--644\relax
\mciteBstWouldAddEndPuncttrue
\mciteSetBstMidEndSepPunct{\mcitedefaultmidpunct}
{\mcitedefaultendpunct}{\mcitedefaultseppunct}\relax
\EndOfBibitem
\bibitem[Cha \emph{et~al.}(2017)Cha, Liu, You, Stephenson, and
  Ulvestad]{Cha2017}
W.~Cha, Y.~Liu, H.~You, G.~B. Stephenson and A.~Ulvestad, \emph{Advanced
  Functional Materials}, 2017, \textbf{27}, 1700331\relax
\mciteBstWouldAddEndPuncttrue
\mciteSetBstMidEndSepPunct{\mcitedefaultmidpunct}
{\mcitedefaultendpunct}{\mcitedefaultseppunct}\relax
\EndOfBibitem
\bibitem[Chen-Wiegart \emph{et~al.}(2017)Chen-Wiegart, Harder, Dunand, and
  McNulty]{Chen-Wiegart2017}
Y.~C.~K. Chen-Wiegart, R.~Harder, D.~C. Dunand and I.~McNulty,
  \emph{Nanoscale}, 2017, \textbf{9}, 5686--5693\relax
\mciteBstWouldAddEndPuncttrue
\mciteSetBstMidEndSepPunct{\mcitedefaultmidpunct}
{\mcitedefaultendpunct}{\mcitedefaultseppunct}\relax
\EndOfBibitem
\bibitem[Hofmann \emph{et~al.}(2017)Hofmann, Tarleton, Harder, Phillips, Ma,
  Clark, Robinson, Abbey, Liu, and Beck]{Hofmann2017a}
F.~Hofmann, E.~Tarleton, R.~J. Harder, N.~W. Phillips, P.~W. Ma, J.~N. Clark,
  I.~K. Robinson, B.~Abbey, W.~Liu and C.~E. Beck, \emph{Scientific Reports},
  2017, \textbf{7}, 1--10\relax
\mciteBstWouldAddEndPuncttrue
\mciteSetBstMidEndSepPunct{\mcitedefaultmidpunct}
{\mcitedefaultendpunct}{\mcitedefaultseppunct}\relax
\EndOfBibitem
\bibitem[Kim \emph{et~al.}(2018)Kim, Chung, Carnis, Kim, Yun, Kang, Cha,
  Cherukara, Maxey, Harder, Sasikumar, {K. R. S. Sankaranarayanan}, Zozulya,
  Sprung, Riu, and Kim]{Kim2018d}
D.~Kim, M.~Chung, J.~Carnis, S.~Kim, K.~Yun, J.~Kang, W.~Cha, M.~J. Cherukara,
  E.~Maxey, R.~Harder, K.~Sasikumar, S.~{K. R. S. Sankaranarayanan},
  A.~Zozulya, M.~Sprung, D.~Riu and H.~Kim, \emph{Nature Communications}, 2018,
  \textbf{9}, 3422\relax
\mciteBstWouldAddEndPuncttrue
\mciteSetBstMidEndSepPunct{\mcitedefaultmidpunct}
{\mcitedefaultendpunct}{\mcitedefaultseppunct}\relax
\EndOfBibitem
\bibitem[Yuan \emph{et~al.}(2019)Yuan, Lee, Cha, Ulvestad, Kim, Abdilla,
  Sturchio, and Fenter]{Yuan2019}
K.~Yuan, S.~S. Lee, W.~Cha, A.~Ulvestad, H.~Kim, B.~Abdilla, N.~C. Sturchio and
  P.~Fenter, \emph{Nature Communications}, 2019, \textbf{10}, 703\relax
\mciteBstWouldAddEndPuncttrue
\mciteSetBstMidEndSepPunct{\mcitedefaultmidpunct}
{\mcitedefaultendpunct}{\mcitedefaultseppunct}\relax
\EndOfBibitem
\bibitem[Miao and Sayre(2000)]{Miao2000b}
J.~Miao and D.~Sayre, \emph{Acta Crystallographica Section A: Foundations of
  Crystallography}, 2000, \textbf{56}, 596--605\relax
\mciteBstWouldAddEndPuncttrue
\mciteSetBstMidEndSepPunct{\mcitedefaultmidpunct}
{\mcitedefaultendpunct}{\mcitedefaultseppunct}\relax
\EndOfBibitem
\bibitem[Robinson \emph{et~al.}(2001)Robinson, Vartanyants, Williams, Pfeifer,
  and Pitney]{Robinson2001}
I.~K. Robinson, I.~A. Vartanyants, G.~J. Williams, M.~A. Pfeifer and J.~A.
  Pitney, \emph{Physical Review Letters}, 2001, \textbf{87}, 1--4\relax
\mciteBstWouldAddEndPuncttrue
\mciteSetBstMidEndSepPunct{\mcitedefaultmidpunct}
{\mcitedefaultendpunct}{\mcitedefaultseppunct}\relax
\EndOfBibitem
\bibitem[Yang \emph{et~al.}(2019)Yang, Phillips, and Hofmann]{Yang2019}
D.~Yang, N.~W. Phillips and F.~Hofmann, \emph{Journal of Synchrotron
  Radiation}, 2019, \textbf{26}, 2055--2063\relax
\mciteBstWouldAddEndPuncttrue
\mciteSetBstMidEndSepPunct{\mcitedefaultmidpunct}
{\mcitedefaultendpunct}{\mcitedefaultseppunct}\relax
\EndOfBibitem
\bibitem[Hofmann \emph{et~al.}(2017)Hofmann, Phillips, Harder, Liu, Clark,
  Robinson, and Abbey]{Hofmann2017b}
F.~Hofmann, N.~W. Phillips, R.~J. Harder, W.~Liu, J.~N. Clark, I.~K. Robinson
  and B.~Abbey, \emph{Journal of Synchrotron Radiation}, 2017, \textbf{24},
  1048--1055\relax
\mciteBstWouldAddEndPuncttrue
\mciteSetBstMidEndSepPunct{\mcitedefaultmidpunct}
{\mcitedefaultendpunct}{\mcitedefaultseppunct}\relax
\EndOfBibitem
\bibitem[Leake \emph{et~al.}(2009)Leake, Newton, Harder, and
  Robinson]{Leake2009}
S.~J. Leake, M.~C. Newton, R.~Harder and I.~K. Robinson, \emph{Optics Express},
  2009, \textbf{17}, 15853\relax
\mciteBstWouldAddEndPuncttrue
\mciteSetBstMidEndSepPunct{\mcitedefaultmidpunct}
{\mcitedefaultendpunct}{\mcitedefaultseppunct}\relax
\EndOfBibitem
\bibitem[Hofmann \emph{et~al.}(2018)Hofmann, Harder, Liu, Liu, Robinson, and
  Zayachuk]{Hofmann2018}
F.~Hofmann, R.~J. Harder, W.~Liu, Y.~Liu, I.~K. Robinson and Y.~Zayachuk,
  \emph{Acta Materialia}, 2018, \textbf{154}, 113--123\relax
\mciteBstWouldAddEndPuncttrue
\mciteSetBstMidEndSepPunct{\mcitedefaultmidpunct}
{\mcitedefaultendpunct}{\mcitedefaultseppunct}\relax
\EndOfBibitem
\bibitem[Hofmann \emph{et~al.}(2020)Hofmann, Phillips, Das, Karamched, Hughes,
  Douglas, Cha, and Liu]{Hofmann2020}
F.~Hofmann, N.~W. Phillips, S.~Das, P.~Karamched, G.~M. Hughes, J.~O. Douglas,
  W.~Cha and W.~Liu, \emph{Physical Review Materials}, 2020, \textbf{4},
  013801\relax
\mciteBstWouldAddEndPuncttrue
\mciteSetBstMidEndSepPunct{\mcitedefaultmidpunct}
{\mcitedefaultendpunct}{\mcitedefaultseppunct}\relax
\EndOfBibitem
\bibitem[Beutier \emph{et~al.}(2013)Beutier, Verdier, Parry, Gilles, Labat,
  Richard, Cornelius, Lory, {Vu Hoang}, Livet, Thomas, and {De
  Boissieu}]{Beutier2013}
G.~Beutier, M.~Verdier, G.~Parry, B.~Gilles, S.~Labat, M.~I. Richard,
  T.~Cornelius, P.~F. Lory, S.~{Vu Hoang}, F.~Livet, O.~Thomas and M.~{De
  Boissieu}, \emph{Thin Solid Films}, 2013, \textbf{530}, 120--124\relax
\mciteBstWouldAddEndPuncttrue
\mciteSetBstMidEndSepPunct{\mcitedefaultmidpunct}
{\mcitedefaultendpunct}{\mcitedefaultseppunct}\relax
\EndOfBibitem
\bibitem[Horng(1968)]{Horng1968}
C.~T. Horng, \emph{PhD thesis}, Missouri University of Science and Technology,
  1968\relax
\mciteBstWouldAddEndPuncttrue
\mciteSetBstMidEndSepPunct{\mcitedefaultmidpunct}
{\mcitedefaultendpunct}{\mcitedefaultseppunct}\relax
\EndOfBibitem
\bibitem[Galanakis \emph{et~al.}(2002)Galanakis, Papanikolaou, and
  Dederichs]{Galanakis2002}
I.~Galanakis, N.~Papanikolaou and P.~H. Dederichs, \emph{Surface Science},
  2002, \textbf{511}, 1--12\relax
\mciteBstWouldAddEndPuncttrue
\mciteSetBstMidEndSepPunct{\mcitedefaultmidpunct}
{\mcitedefaultendpunct}{\mcitedefaultseppunct}\relax
\EndOfBibitem
\bibitem[Larijani \emph{et~al.}(2011)Larijani, Yari, Afshar, Jafarian, and
  Eshghabadi]{Larijani2011}
M.~M. Larijani, M.~Yari, A.~Afshar, M.~Jafarian and M.~Eshghabadi,
  \emph{Journal of Alloys and Compounds}, 2011, \textbf{509}, 7400--7404\relax
\mciteBstWouldAddEndPuncttrue
\mciteSetBstMidEndSepPunct{\mcitedefaultmidpunct}
{\mcitedefaultendpunct}{\mcitedefaultseppunct}\relax
\EndOfBibitem
\bibitem[L{\"{u}}ttge \emph{et~al.}(2013)L{\"{u}}ttge, Arvidson, and
  Fischer]{Luttge2013}
A.~L{\"{u}}ttge, R.~S. Arvidson and C.~Fischer, \emph{Elements}, 2013,
  \textbf{9}, 183--188\relax
\mciteBstWouldAddEndPuncttrue
\mciteSetBstMidEndSepPunct{\mcitedefaultmidpunct}
{\mcitedefaultendpunct}{\mcitedefaultseppunct}\relax
\EndOfBibitem
\bibitem[Briese \emph{et~al.}(2017)Briese, Arvidson, and Luttge]{Briese2017}
L.~Briese, R.~S. Arvidson and A.~Luttge, \emph{Geochimica et Cosmochimica
  Acta}, 2017, \textbf{212}, 167--175\relax
\mciteBstWouldAddEndPuncttrue
\mciteSetBstMidEndSepPunct{\mcitedefaultmidpunct}
{\mcitedefaultendpunct}{\mcitedefaultseppunct}\relax
\EndOfBibitem
\bibitem[Vi{\~{n}}es \emph{et~al.}(2010)Vi{\~{n}}es, Lykhach, Staudt, Lorenz,
  Papp, Steinr{\"{u}}ck, Libuda, Neyman, and G{\"{o}}rling]{Vines2010}
F.~Vi{\~{n}}es, Y.~Lykhach, T.~Staudt, M.~P. Lorenz, C.~Papp, H.~P.
  Steinr{\"{u}}ck, J.~Libuda, K.~M. Neyman and A.~G{\"{o}}rling,
  \emph{Chemistry - A European Journal}, 2010, \textbf{16}, 6530--6539\relax
\mciteBstWouldAddEndPuncttrue
\mciteSetBstMidEndSepPunct{\mcitedefaultmidpunct}
{\mcitedefaultendpunct}{\mcitedefaultseppunct}\relax
\EndOfBibitem
\bibitem[Clark \emph{et~al.}(2015)Clark, Ihli, Schenk, Kim, Kulak, Campbell,
  Nisbet, Meldrum, and Robinson]{Clark2015}
J.~N. Clark, J.~Ihli, A.~S. Schenk, Y.~Y. Kim, A.~N. Kulak, J.~M. Campbell,
  G.~Nisbet, F.~C. Meldrum and I.~K. Robinson, \emph{Nature Materials}, 2015,
  \textbf{14}, 780--784\relax
\mciteBstWouldAddEndPuncttrue
\mciteSetBstMidEndSepPunct{\mcitedefaultmidpunct}
{\mcitedefaultendpunct}{\mcitedefaultseppunct}\relax
\EndOfBibitem
\bibitem[Mesu \emph{et~al.}(2006)Mesu, Beale, {De Groot}, and
  Weckhuysen]{Mesu2006}
J.~G. Mesu, A.~M. Beale, F.~M. {De Groot} and B.~M. Weckhuysen, \emph{Journal
  of Physical Chemistry B}, 2006, \textbf{110}, 17671--17677\relax
\mciteBstWouldAddEndPuncttrue
\mciteSetBstMidEndSepPunct{\mcitedefaultmidpunct}
{\mcitedefaultendpunct}{\mcitedefaultseppunct}\relax
\EndOfBibitem
\bibitem[Chen \emph{et~al.}(2007)Chen, Miao, Wang, and Lee]{Chen2007}
C.~C. Chen, J.~Miao, C.~W. Wang and T.~K. Lee, \emph{Physical Review B -
  Condensed Matter and Materials Physics}, 2007, \textbf{76}, 064113\relax
\mciteBstWouldAddEndPuncttrue
\mciteSetBstMidEndSepPunct{\mcitedefaultmidpunct}
{\mcitedefaultendpunct}{\mcitedefaultseppunct}\relax
\EndOfBibitem
\bibitem[Marchesini \emph{et~al.}(2003)Marchesini, He, Chapman, Hau-Riege, Noy,
  Howells, Weierstall, and Spence]{Marchesini2003}
S.~Marchesini, H.~He, N.~Chapman, P.~Hau-Riege, A.~Noy, R.~Howells,
  U.~Weierstall and H.~Spence, \emph{Physical Review B - Condensed Matter and
  Materials Physics}, 2003, \textbf{68}, 1401011--1401014\relax
\mciteBstWouldAddEndPuncttrue
\mciteSetBstMidEndSepPunct{\mcitedefaultmidpunct}
{\mcitedefaultendpunct}{\mcitedefaultseppunct}\relax
\EndOfBibitem
\bibitem[Ulvestad \emph{et~al.}(2017)Ulvestad, Nashed, Beutier, Verdier,
  Hruszkewycz, and Dupraz]{Ulvestad2017b}
A.~Ulvestad, Y.~Nashed, G.~Beutier, M.~Verdier, S.~O. Hruszkewycz and
  M.~Dupraz, \emph{Scientific Reports}, 2017, \textbf{7}, 9920\relax
\mciteBstWouldAddEndPuncttrue
\mciteSetBstMidEndSepPunct{\mcitedefaultmidpunct}
{\mcitedefaultendpunct}{\mcitedefaultseppunct}\relax
\EndOfBibitem
\bibitem[Clark \emph{et~al.}(2012)Clark, Huang, Harder, and
  Robinson]{Clark2012}
J.~Clark, X.~Huang, R.~Harder and I.~Robinson, \emph{Nature Communications},
  2012, \textbf{3}, 993\relax
\mciteBstWouldAddEndPuncttrue
\mciteSetBstMidEndSepPunct{\mcitedefaultmidpunct}
{\mcitedefaultendpunct}{\mcitedefaultseppunct}\relax
\EndOfBibitem
\bibitem[Richardson(1972)]{Richardson1972}
W.~H. Richardson, \emph{Journal of the Optical Society of America}, 1972,
  \textbf{62}, 55--59\relax
\mciteBstWouldAddEndPuncttrue
\mciteSetBstMidEndSepPunct{\mcitedefaultmidpunct}
{\mcitedefaultendpunct}{\mcitedefaultseppunct}\relax
\EndOfBibitem
\bibitem[Constantinescu and Korsunsky(2007)]{Constantinescu2008}
A.~Constantinescu and A.~Korsunsky, \emph{{Elasticity with
  MATHEMATICA{\textregistered}: An introduction to continuum mechanics and
  linear elasticity}}, Cambridge University Press, Cambridge, 2007, vol.
  9780521842, pp. 1--255\relax
\mciteBstWouldAddEndPuncttrue
\mciteSetBstMidEndSepPunct{\mcitedefaultmidpunct}
{\mcitedefaultendpunct}{\mcitedefaultseppunct}\relax
\EndOfBibitem
\bibitem[Newton \emph{et~al.}(2010)Newton, Leake, Harder, and
  Robinson]{Newton2010}
M.~C. Newton, S.~J. Leake, R.~Harder and I.~K. Robinson, \emph{Nature
  Materials}, 2010, \textbf{9}, 120--124\relax
\mciteBstWouldAddEndPuncttrue
\mciteSetBstMidEndSepPunct{\mcitedefaultmidpunct}
{\mcitedefaultendpunct}{\mcitedefaultseppunct}\relax
\EndOfBibitem
\bibitem[Richards(2015)]{Richards2015}
A.~Richards, \emph{{University of Oxford Advanced Research Computing}}, 2015,
  \url{https://zenodo.org/record/22558}\relax
\mciteBstWouldAddEndPuncttrue
\mciteSetBstMidEndSepPunct{\mcitedefaultmidpunct}
{\mcitedefaultendpunct}{\mcitedefaultseppunct}\relax
\EndOfBibitem
\end{mcitethebibliography}
\bibliographystyle{rsc} 

\end{document}